\documentclass[epj,nopacs]{svjour}
\usepackage{graphics}
\begin{document}
\title{Dimensionally regulated
       one$-$loop box scalar integrals
       with massless internal lines}
\author{G. Duplan\v ci\' c\thanks{gorand@thphys.irb.hr} 
\and B. Ni\v zi\' c\thanks{nizic@thphys.irb.hr}% etc
}                     % Do not remove
\institute{Theoretical Physics Division, Rudjer Bo\v{s}kovi\'{c} Institute, 
        P.O. Box 180, HR-10002 Zagreb, Croatia}
\date{Received: date / Revised version: date}
\abstract{
Using the Feynman parameter method, we have calculated 
in an elegant manner a set
of one$-$loop box scalar integrals with massless internal lines,
but containing 0, 1, 2, or 3 external massive lines. To treat
IR divergences (both soft and collinear), the dimensional
regularization method has been employed. The results for these integrals, 
which appear in the process of evaluating one$-$loop $(N\ge 5)-$point
integrals and in subdiagrams in QCD loop calculations, have been 
obtained for arbitrary values of the relevant kinematic variables and
presented in a simple and compact form.
\PACS{
      {PACS-key}{discribing text of that key}   \and
      {PACS-key}{discribing text of that key}
     } % end of PACS codes
} %end of abstract
\maketitle
\section{Introduction}
\label{intro}

Scattering processes are one of the most important sources of information
on short$-$distance physics and have played a vital role in establishing
the fundamental interactions of nature.
In testing various aspects of QCD, the scattering processes in which
the total number of particles in the initial and final states is
$N\ge5$ (like 2 $\rightarrow$ 3, 2 $\rightarrow$ 4, etc.) are becoming
increasingly important.

The techniques for calculating the tree level amplitudes involving
a large number of particles in the final state are well established \cite{tree}.
Owing to the well$-$known fact that the LO predictions in
perturbative QCD do not have much predictive power, the inclusion of
higher$-$order corrections is essential for many reasons.
In general, higher- order corrections have a stabilizing effect
reducing the dependence of the LO predictions on the
renormalization and factorization scales and the
renormalization scheme.
Therefore, to achieve a complete confrontation between theoretical
predictions and experimental data, it is very important to know the
size of radiative corrections to the LO predictions.

Obtaining radiative corrections requires evaluation of one$-$loop
integrals which arise from a Feynman diagramatic approach.
The case of massless internal lines is of special interest,
because we often deal with either really massless particles
(gluons) or particles whose masses can be neglected in
high$-$energy processes (quarks).
The main technical difficulty in obtaining the NLO corrections
consists in the treatment of the occurring $N$$-$point tensor
and scalar integrals with massless internal lines.
In QCD, tensor integrals appear in which the $N$$-$point
integral may contain up to $N$ powers of the loop momentum in the
numerator of the integrand.
Since these integrals contain IR
divergences, they need to be calculated in an arbitrary number of dimensions
and the standard methods of \cite{old} cannot be directly applied.

Various approaches have been proposed for reducing the dimensionally regulated
$(N\ge 5)$$-$point integrals to a linear combination of $N-$ and lower$-$point
scalar integrals
multiplied by tensor structures made from the metric tensor
$g^{\mu \nu}$ and external momenta \cite{davidicev,dixon,tarasov,binoth,camp}.

It has also been shown that the general $(N\ge 5)$$-$point scalar one$-$loop 
integral can be recursively represented as a 
cyclically symmetric combination of $(N-1)$$-$point
integrals, provided the external momenta are kept in
four dimensions \cite{dixon,tarasov,binoth}.
Consequently, all scalar integrals occurring in the 
computation of an arbitrary one$-$loop $(N \ge 5)$$-$point integral
with massless internal lines
can be reduced to a sum over a set of basic scalar box
$(N=4)$ integrals with rational coefficients 
depending on the external momenta and the dimensionality of space$-$time.
This set of diagrams includes IR divergent box integrals with 
massless internal lines but containing 0, 1, 2, and 3 external masses
and the IR finite box integral with four external masses.

The IR finite box integral has been evaluated in Ref. \cite{hooft}, and
written in a more compact form in Ref. \cite{denner}.
The results for the IR divergent box integrals
with no external masses and with one external mass have been obtained in
Refs. \cite{fabricius,papa}. All IR divergent box integrals have been 
considered in Ref. \cite{bern}, 
using the partial differential equation technique. 
However, the results obtained for the
integrals containing two and three external masses are strictly  correct
only in the Euclidean region where all 
relevant kinematical variables are negative.

Being very nontrivial, impossible to check numerically, and
of fundamental importance for one$-$loop calculations in
perturbative QCD with massless quarks, it is absolutely essential that these
integrals should be evaluated and the results of Ref. \cite{bern}
should be checked using independent techniques.

In this paper we recalculate the IR one$-$loop box scalar
integrals in an elegant manner using dimensional regularisation and
the Feynman parameter method, and give results in a simple and compact form.
A characteristic feature of our calculation is that the
causal ${\rm i}\epsilon$ has been systematically kept throughout the
calculation, so that the results obtained are valid for
arbitrary values of the relevant kinematic variables.

The paper is organized as follows.
Section 2 is devoted to introducing the notation and to some
preliminary considerations.
In Sec. 3, using the Feynman parameter method and  
the dimensional regularization method, we evaluate the IR divergent
one$-$loop box Feynman integrals with massless internal
lines but containing 0, 1, 2, and 3 massive external lines, and compare
our results with the corresponding ones obtained in Ref. \cite{bern}.
Section 4 is devoted to some concluding remarks.
An analytical proof of the equivalence of our results
to those obtained in Ref. \cite{bern} for Euclidean kinematics 
is given in Appendix A.
For the reader's convenience, in Appendix B we present the
closed$-$form expressions for the IR divergent one$-$loop scalar box
integrals evaluated in this paper, i.e., with all poles in
${\epsilon}_{IR}=D/2-2$ manifest, and with all functions of the kinematic
variables expressed in terms of logarithms and dilogarithms.

\section{Preliminaries}
\label{sec:1}

\begin{figure}
% \resizebox{0.75\textwidth}{!}{%
%  \includegraphics{fig1.ps}
%}
 \resizebox{0.5\textwidth}{!}{%
  \includegraphics{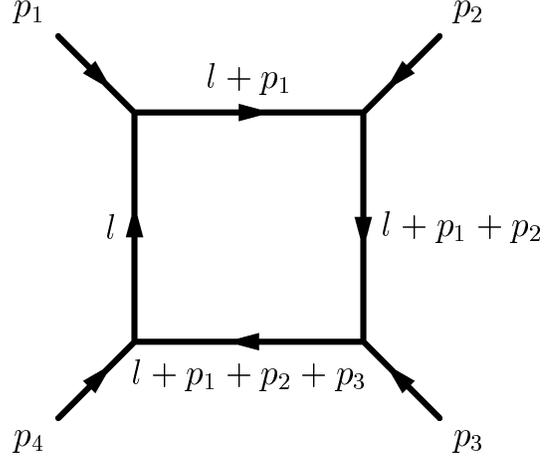}
}
 \caption{Basic one-loop box diagram.}
 \label{f:fig1}
\end{figure}

The massless scalar one$-$loop box integral in
$D$$-$di\-men\-si\-o\-nal
space$-$time is given by
\begin{equation}
I_{4}(p_1,p_2,p_3,p_4)=({\mu}^2)^{2-D/2}
\int \frac{{\rm d}^{D}l}{(2\pi)^{D}}\frac{1}
{A_1 A_2 A_3 A_4}~,\label{eq:f1}
\end{equation}
where $p_i$, $i$=1,2,3,4 are the external momenta, $l$ is the loop
momentum, and $\mu$ is the usual dimensional regularization scale.
As indicated in Fig. \ref{f:fig1}, all external momenta are taken to
 be incoming,
so that the massless propagators have the form
\begin{eqnarray}
A_{1} &=& l^{2}+{\rm i}\epsilon \,,\nonumber \\
A_{2} &=& (l+p_{1})^{2}+{\rm i}\epsilon \,,\nonumber \\
A_{3} &=& (l+p_{1}+p_{2})^{2}+{\rm i}\epsilon \,,\nonumber \\
A_{4} &=& (l+p_{1}+p_{2}+p_{3})^{2}+{\rm i}\epsilon \,.\label{eq:f2} 
\end{eqnarray}

\noindent
Combining the denominators with the help of the Feynman 
parametrization formula
\begin{eqnarray}
\lefteqn{\frac{1}{A_1 A_2 A_3 A_4}=} \label{eq:f3}\\
& &=\int_{0}^{1}\!\!{\rm d}x_1{\rm d}x_2 
{\rm d}x_3{\rm d}x_4 
\frac{3!\,\delta(x_1+x_2+x_3+x_4-1)}{(x_1 A_1+x_2 A_2+x_3 A_3+x_4 A_4)^4},
\nonumber
\end{eqnarray}
performing the $D$$-$dimensional loop momentum integration using
\begin{equation}
\int \frac{{\rm d}^{D}l}{(2\pi)^{D}}
\frac{1}{(l^{2}-M^{2}+{\rm i}\epsilon)^4}=
\frac{{\rm i}}{(4\pi)^{D/2}}\frac{\Gamma(4-D/2)}
{3!\,(M^2-{\rm i}\epsilon)^{4-D/2}}~,\label{eq:f4}
\end{equation}
introducing the external "masses"
\begin{equation}
p_i^2=m_i^2~~(i=1,2,3,4)\, ,\label{eq:f5}
\end{equation}
and the Mandelstam variables
\begin{equation}
s=(p_1+p_2)^2, ~~~t=(p_2+p_3)^2, \label{eq:f6}
\end{equation}
one readily finds that the scalar integral in (\ref{eq:f1})
can be written in the form
\begin{eqnarray}
\lefteqn{I_4(s,t;m_1^2,m_2^2,m_3^2,m_4^2)= 
\frac{{\rm i}}{(4\pi)^2}
\frac{\Gamma (4-D/2)}
{(4\pi {\mu}^2)^{D/2-2}}}\nonumber \\
& &\times \int_0^1{\rm d}x_1 {\rm d}x_2 {\rm d}x_3 {\rm d}x_4 
\;\delta(x_1+x_2+x_3+x_4-1)
\nonumber \\
& &\times \left(-x_1x_3\;s-x_2x_4\;t-x_1x_2\;m^2_1\right . \nonumber \\
& &{}\left .-x_2x_3\;m^2_2
-x_3x_4\;m^2_3-x_1x_4\;m^2_4-{\rm i}\epsilon\right)^{D/2-4}. \label{eq:f7}
\end{eqnarray}
This is the basic four$-$point "scalar" parametric integral, serving as a
starting point for our further considerations. 

Depending on the number of the external massless lines, we distinguish
six special cases of this integral.
Following the notation of Ref. \cite{bern}, we denote these integrals by
\begin{eqnarray}
I_4^{4m} &\equiv & I_4(s,t;m_1^2,m_2^2,m_3^2,m_4^2), \label{eq:f10} \\
I_4^{3m} &\equiv & I_4(s,t;0,m_2^2,m_3^2,m_4^2), \label{eq:f11} \\
I_4^{2mh} &\equiv & I_4(s,t;0,0,m_3^2,m_4^2), \label{eq:f12} \\
I_4^{2me} &\equiv & I_4(s,t;0,m_2^2,0,m_4^2), \label{eq:f13} \\
I_4^{1m} &\equiv & I_4(s,t;0,0,0,m_4^2), \label{eq:f14} \\
I_4^{0m} &\equiv & I_4(s,t;0,0,0,0,), \label{eq:f15} 
\end{eqnarray}
and refer to them as the four$-$mass scalar box integral,
the three$-$mass box integral, the "hard" two$-$mass box integral,
the "easy" two$-$mass box integral, the one$-$mass box integral,
and the massless box integral, respectively.

These six box integrals constitute the fundamental
set of integrals, in the sense that an arbitrary one$-$loop $N(\ge 5)-$point
integral with massless internal lines can be represented in a unique way as
a linear combination of these integrals with the coefficients being
rational functions of the relevant kinematic variables and the number of
space$-$di\-men\-si\-ons $D$.

\begin{figure}
  \resizebox{0.5\textwidth}{!}{%
  \includegraphics{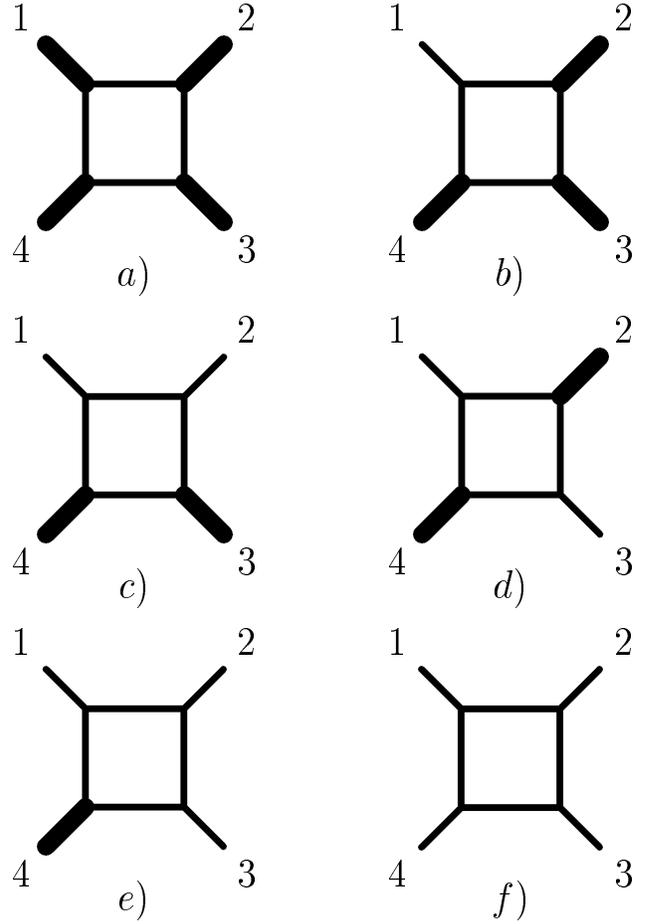}
}
 \caption{One-loop box diagrams with massless internal lines but containing 0,
1, 2, 3, and 4 external massive lines; a) the four$-$mass box diagram; b) the
three$-$mass box diagram; c) the two$-$mass box diagram with external masses at
adjacent corners (legs 3 and 4); d) the two$-$mass box diagram with external
masses at diagonally opposite corners (legs 2 and 4); e) the one$-$mass box
diagram; f) the massless box diagram. Thick lines designate the massive external
lines.}
 \label{f:fig2}
\end{figure}

These integrals arise from the Feynman diagrams depicted in Fig. \ref{f:fig2},
formally corresponding to the scalar massless ${\sl \Phi}^3$ theory.
The thick lines in these diagrams denote the massive (off$-$shell)
external lines. As it is seen from Fig. \ref{f:fig2}, there are two distinct
configurations related to the case when two external lines are
massless: the adjacent box diagram ($m_1^2=m_2^2=0$)
and the opposite box diagram ($m_1^2=m_3^2=0$). They correspond to the
hard and easy two$-$mass box integrals, respectively.

When evaluating the diagrams of Fig. \ref{f:fig2}, one comes across IR singularities
(both collinear and soft). Let us recall the circumstances under which
IR singularities appear in a Feynman diagram.
When an on$-$shell quark of momentum $p$ emits a
gluon of momentum $k$, then IR singularities appear as a result of
vanishing of the quark propagator. If the quark is massless, this can
happen when either $p$ and $k$ are collinear ($k\parallel p$,
collinear singularity) or when the gluon momentum vanishes
($k\rightarrow 0$, soft singularity). Thus, a Feynman diagram with all
particles massless will have
 soft singularity if it contains an internal gluon line attached to
two on$-$shell external quark lines. On the other hand, a diagram will contain
 collinear singularity if it has an internal gluon line attached to an
on$-$shell external quark line. It follows then that a diagram containing soft
singularity contains two collinear singularities at the same time, i.e.,
soft and collinear singularities overlap.

In view of what has been said above, we conclude that the integral
$I_4^{4m}$ is $IR$ finite, and can be calculated in $D=4$ space$-$time
dimensions, while the rest of integrals contain IR divergence
($I_4^{3m}$ and $I_4^{2me}$ collinear divergence,
$I_4^{2mh}$, $I_4^{1m}$, and $I_4^{0m}$ both  collinear and soft
divergence), and as such have to be evaluated in
$D=4+2{\varepsilon}_{IR}~~({\varepsilon}_{IR}>0)$ dimensions.

For arbitrary $D$, the integrals (\ref{eq:f11}$-$\ref{eq:f15}) cannot
be expressed by elementary
functions, but we know that the expressions corresponding to these
diagrams expanded in powers of ${\varepsilon}_{IR}$ are of the generic form
\[
\sim \frac{A}{\varepsilon_{IR}^2}
+\frac{B}{\varepsilon_{IR}}
+C
+\mathcal{O}({\epsilon}_{IR})~,
\]
with the coefficients $A,~B,~C$ being complex functions of the
kinematic invariants.
The $1/{\varepsilon}_{IR}$ poles express the IR divergence.

As stated in the Introduction, the IR divergent integrals $I_4^K$, $K \in 
{\{} 3m, 2mh, 2me, 1m, 0m {\}} $, defined through Eqs. (\ref{eq:f7}) and 
(\ref{eq:f11})$-$(\ref{eq:f15})
, have been evaluated in Ref. \cite{bern} using the partial differential
equation technique. 
After experimenting with various ways to independently derive the results of
Ref. \cite{bern},
we have found that the Feynman parameter method appears to be the
most straightforward and satisfactory approach.

\section{Calculation and results}
\label{sec:2}

Using the Feynman parameter method and the method of dimensional
regularization, in this section we evaluate the IR divergent scalar
one$-$loop box integrals
$I_4^K$, $K\in \{ 3m,2mh,2me,1m,0m\} $.
To accomplish that, we start by considering the most complicated
of these integrals, namely, the three$-$mass box integral $I_4^{3m}$.
We show that, paying due attention to the fact that the
limit of taking a mass to zero does not necessarily commute with the
${\varepsilon}_{IR}$ expansion of dimensional regularization, the
results obtained at the intermediate steps of the calculation of the
integral $I_4^{3m}$ can be used to obtain the results for the rest
of the above integrals. A characteristic feature of our calculation is that we
keep the causal ${\rm i}\epsilon$ systematically through the calculation, so that
the results we obtain are valid for arbitrary values of the relevant kinematic
variables.

Let us then start with the three$-$mass box integral $I_4^{3m}$.
By setting $m_1^2=0$ and $D=4+2{\varepsilon}_{IR}~~({\varepsilon}_{IR}>0)$
in Eq. (\ref{eq:f7}) and eliminating the $\delta -$function
by performing the $x_4$ integration the integral $I_4^{3m}$ becomes
\begin{eqnarray}
I_4^{3m}&=&
\frac{{\rm i}}{(4\pi)^2}
\frac{\Gamma (2-{\varepsilon}_{IR})}
{(4\pi {\mu}^2)^{{\varepsilon}_{IR}}}
\,
\int_0^1{\rm d}x_1 \int_0^{1-x_1}\!\!\!{\rm d}x_2 \int_0^{1-x_1-x_2}
\!\!\!\!{\rm d}x_3
\nonumber \\
& &\times \left[ \,-x_1x_3\;s-x_2x_3\;m^2_2
-\left( 1-x_1-x_2-x_3 \right) \right. \nonumber \\
& &\times \left. \left( x_1\;m^2_4+x_2\;t+x_3\;m^2_3\right) 
-{\rm i}\epsilon \,\right]^{{\varepsilon}_{IR}-2}. \label{eq:fgo}
\end{eqnarray}
It is a well$-$known fact that the appropriate choice of Feynman parameters is 
in practice a critical ingredient in enabling one to evaluate a complicated
Feynman integral analytically. There does not appear to be any simple 
formula for choosing an optimal set of Feynman parameters for a given
diagram, as there is generally an enormous set of possibilities.

To proceed with the evaluation of the integral $I_4^{3m}$, the most suitable set of
Feynman parameters turns out to be given by \cite{karplus}
\begin{eqnarray}
x_1 &=&(1-x)\, (1-y),\nonumber \\
x_2 &=& x\, (1-y), \nonumber \\
x_3 &=& y\, z. \label{eq:f8}  
\end{eqnarray}
The Jacobian corresponding to this transformation of the integration variables 
is $y(1-y)$.
Written in terms of the new variables, 
the integral (\ref{eq:fgo}) takes the form
\begin{eqnarray}
\lefteqn{I_4^{3m} = 
\frac{{\rm i}}{(4\pi)^2}
\frac{\Gamma (2-{\varepsilon}_{IR})}
{(4\pi {\mu}^2)^{{\varepsilon_{IR}}}}
\int_{0}^{1}\!\!\!{\rm d}x\, {\rm d}y\, {\rm d}z\:
\, y(1-y)  \Big{\{ } \!-y(1-y)}\, \nonumber \\
& &
\times \Big[\,
(1-x)z\, s+(1-x)(1-z)\, m_{4}^{2}+xz\, m_{2}^{2}+x(1-z)\, t\,\Big]
\nonumber \\
& & {}-z(1-z) y^2\, m_{3}^{2}
-{\rm i}\epsilon\,\Big{\} }^{\varepsilon_{IR}-2}.\label{eq:f16}
\end{eqnarray}
As it is seen from Eq. (\ref{eq:f16}), the integration over $x$ is elementary and
is readily performed. The resulting expression is
\begin{eqnarray}
I_4^{3m} &=&
\frac{\kappa}{2} 
\int_{0}^{1}{\rm d}y\, {\rm d}z\:
\frac{1}{z\,(s-m_2^2)+(1-z)\,(m_4^2-t)}
\nonumber \\
& &
\times \Big{\{ } \, \Big[
-y(1-y)\Big( z\, s+(1-z)\, m_4^2\Big)\nonumber \\
& &{}-z(1-z) y^2\,
m_{3}^{2}-{\rm i}\epsilon 
\Big]^{\varepsilon_{IR}-1}
\nonumber \\
& & -\Big[
-y(1-y)\Big( z\, m_2^2+(1-z)\, t\Big)\nonumber \\
& &{}-z(1-z) y^2\,
m_{3}^{2}-{\rm i}\epsilon 
\Big]^{\varepsilon_{IR}-1}\, 
\Big{\} },\label{eq:f17}
\end{eqnarray}
where we have introduced the abbreviation
\begin{equation}
\kappa=
\frac{{\rm i}}{(4\pi)^2}
\frac{2\,\Gamma (1-{\varepsilon}_{IR})}
{(4\pi {\mu}^2)^{{\varepsilon_{IR}}}}~\cdot \label{eq:f18}
\end{equation}
Next, by pulling out the factor $y^{\varepsilon_{IR}-1}$ from both terms in the
curly brackets (which is legitimate since $y$ is positive in the integration
region and the sign of $\epsilon(>0)$ does not change), and making use
of the following relation:
\begin{eqnarray}
(a+b-{\rm i}\epsilon)^{\varepsilon_{IR}-1} &=& 
\Big(\frac{a}{b-{\rm i}\epsilon}+1 \Big)^{\varepsilon_{IR}-1} 
(b-{\rm i}\epsilon)^{\varepsilon_{IR}-1}\nonumber \\  
& &(a, b\in \mathbf{R}),\label{eq:f19}
\end{eqnarray}
(which is not self$-$evident 
owing to the fact that $\varepsilon_{IR}$ is not an integer)
leads to the integral of the form
\begin{eqnarray}
\lefteqn{I_4^{3m}=
\frac{\kappa}{2} 
\int_{0}^{1}{\rm d}z\:
\frac{1}{z\,(s-m_2^2)+(1-z)\,(m_4^2-t)}}
\nonumber \\
& &
\times
\Bigg \{
\,
\Big(-z\,s-(1-z)\, m_4^2-{\rm i}\epsilon\Big)^{\varepsilon_{IR}-1}
\int_0^1 {\rm d}y\,y^{\varepsilon_{IR}-1}\nonumber \\
& &
\times
\left [1-y
\left (1-\frac{z(1-z) \, m_{3}^{2}+{\rm i}\epsilon}
{ z\, s+(1-z)\, m_4^2+{\rm i}\epsilon}
\right )
\right ]^{\varepsilon_{IR}-1}
\nonumber \\
& &
-\Big( -z\, m_2^2-(1-z)\,t-{\rm i}\epsilon\Big)^{\varepsilon_{IR}-1}
\int_0^1
{\rm d}y\,y^{\varepsilon_{IR}-1}\nonumber \\
& &
\times
\left [1-y
\left (1-\frac{z(1-z) \, m_{3}^{2}+{\rm i}\epsilon}
{ z\, m_2^2+(1-z)\, t+{\rm i}\epsilon}\right )
\right ]^{\varepsilon_{IR}-1}
\Bigg \}. \label{eq:f20}
\end{eqnarray}
By noticing that the integral over $y$ stands for the Euler integral representation
of the hypergeometric function
\begin{eqnarray}
_{2}F_{1}(a,b;c;z) &=&
\frac{\Gamma(c)}{\Gamma(b)\Gamma(c-b)}\int_{0}^{1}{\rm d}t
\frac{t^{b-1}(1-t)^{c-b-1}}{(1-t\,z)^{a}}~,\nonumber \\
& &\mbox{Re}~c>\mbox{Re}~b>0\;;\;\mid \mbox{arg}(1-z)\mid <\pi, \label{eq:f21} 
\end{eqnarray}
we obtain the result
\begin{eqnarray}
\lefteqn{I_4^{3m}= 
\frac{\kappa}{2\,{\varepsilon}_{IR}}
\int_{0}^{1}{\rm d}z\:
\frac{1}{z\,(s-m_2^2)+(1-z)\,(m_4^2-t)}}\nonumber \\
& & \times \Bigg [ \Big( -z\, s-(1-z)\, m_4^2-
{\rm i}\epsilon\Big)^{\varepsilon_{IR}-1}\nonumber \\
& &\times{}_{2}F_{1}\!\!\left (1-\varepsilon_{IR}, \varepsilon_{IR};
1+\varepsilon_{IR}; 1-\frac{z(1-z)\, m_{3}^{2}+{\rm i}\epsilon}
{ z\, s+(1-z)\,m_4^2\, +{\rm i}\epsilon}\right ) \nonumber \\
& &
-\Big( -z\, m_2^2-(1-z)\, t-{\rm i}\epsilon\Big)^{\varepsilon_{IR}-1}
\nonumber \\
& &\times{}_{2}F_{1}\!\!\left (1-\varepsilon_{IR}, \varepsilon_{IR};
1+\varepsilon_{IR};1-\frac{z(1-z)\, m_{3}^{2}+{\rm i}\epsilon}
{ z\, m_2^2+(1-z)\, t+{\rm i}\epsilon}\right ) 
\!\!\Bigg ].\nonumber \\ & &\label{eq:f22}
\end{eqnarray}
Next, using the identity (partial fraction decomposition)
\begin{equation}
\frac{1}{a\,z+b}~\frac{1}{c\,z+d}=\frac{1}{a d-b c}~\left (
\frac{a}{a\,z+b}-\frac{c}{c\,z+d}\right )~,\label{eq:f23} 
\end{equation}
and performing a few simple rearrangements, we find that the integral
under consideration can be written in the following form:
\begin{equation}
I_4^{3m}=
\frac{\kappa}{st-m_2^2m_4^2}
(P^{3m}+Q^{3m})~,\label{eq:f24} 
\end{equation}
where
\begin{eqnarray}
\lefteqn{P^{3m} =\frac{1}{2\,\varepsilon_{IR}}\Bigg[
(m_2^2-t) 
\!\int_{0}^{1}\!\!\!\!{\rm d}z\:
[-z\,m_2^2-(1-z)\,t-{\rm i}\epsilon]^{\varepsilon_{IR}-1}}
\nonumber \\
& &
\times {}_{2}F_{1}\!\!
\left (1-\varepsilon_{IR}, \varepsilon_{IR}; 1+\varepsilon_{IR};
1-\frac{z(1-z)\, m_{3}^{2}+{\rm i}\epsilon}
{ z\, m_2^2+(1-z)\, t+{\rm i}\epsilon}\right ) \nonumber \\
& &
+(m_4^2-s)
\!\int_{0}^{1}\!\!\!{\rm d}z\:
[-z\,s-(1-z)\,m_4^2-{\rm i}\epsilon]^{\varepsilon_{IR}-1}
\nonumber \\
& &
\times {}_{2}F_{1}\!\!
\left (1-\varepsilon_{IR}, \varepsilon_{IR}; 1+\varepsilon_{IR};
 1-\frac{z(1-z)\, m_{3}^{2}+{\rm i}\epsilon}
{ z\, s+(1-z)\, m_4^2+{\rm i}\epsilon}\right )\!\!\Bigg]
,\nonumber \\ & &\label{eq:f27}
\end{eqnarray}
and
\begin{eqnarray}
\lefteqn{Q^{3m}=\frac{1}{2\,\varepsilon_{IR}}\int_{0}^{1}{\rm d}z\:
\frac{1}{z-z_0}}
\nonumber \\
& &
\times
 \Bigg[(-z\,m_2^2-(1-z)\,t-{\rm i}\epsilon)^{\varepsilon_{IR}} \nonumber \\
& &
\times{}_{2}F_{1}\!\!\left (1-\varepsilon_{IR}, \varepsilon_{IR}; 1+\varepsilon_{IR};
1-\frac{z(1-z)\, m_{3}^{2}+{\rm i}\epsilon}
{ z\, m_2^2+(1-z)\, t+{\rm i}\epsilon}\right )\nonumber \\
& &
-(-z\,s-(1-z)\,m_4^2-{\rm i}\epsilon)^{\varepsilon_{IR}}\nonumber \\
& &
\times{}_{2}F_{1}\!\!\left (1-\varepsilon_{IR}, \varepsilon_{IR};
1+\varepsilon_{IR}; 1-\frac{z(1-z)\, m_{3}^{2}+{\rm i}\epsilon}
{ z\, s+(1-z)\, m_4^2+{\rm i}\epsilon}\right ) \!\!\Bigg],\nonumber \\
 \label{eq:f25}
\end{eqnarray}
with
\begin{equation}
z_0=
\frac{t-m_4^2}{s+t-m_2^2-m_4^2}~.\label{eq:f26}
\end{equation}
In view of (\ref{eq:f24}), it is clear that the integrals 
$I_4^K$ can be written in the form
\begin{eqnarray}
I_{\displaystyle 4}^{K}&=&
\kappa
~g^K
(P^{K}+Q^{K})~,\nonumber \\
& &
K\in \{3m,2mh,2me,1m,0m\} , \label{eq:f28}
\end{eqnarray}
with
\begin{equation}
g^{3m}=g^{2me}=
\frac{1}{st-m_2^2m_4^2}~,~~~~~
g^{2mh}=g^{1m}=g^{0m}=
\frac{1}{st}~\cdot\label{eq:f29}
\end{equation}
Not being able to perform the remaining integrations in
Eqs.  (\ref{eq:f25}) and (\ref{eq:f27}) analytically, 
we proceed by expanding the
integrands in power series in ${\varepsilon}_{IR}$ and by term by term
integration. We shall see below that all divergences of the integral 
$I_4^K$ are contained in $P^K$, while $Q^K$ is completely finite.

It is clear from Eqs. (\ref{eq:f18}) and (\ref{eq:f24}) that to obtain the
values of the integrals $I_4^K$, $K\in \{ 3m,2mh,2me,1m,0m\} $ to order
$\mathcal{O} (\varepsilon_{IR}^0)$, the evaluation of the integrals $P^K$ and
$Q^K$ should be made to the same order.

It turns out, however, that the number
of integrals that really need to be evaluted reduces to just four,
namely, $P^{3m}$, given by (\ref{eq:f27}), the integrals
$P^{2mh}$ and $P^{2me}$, obtained by setting $m_2^2=0$,
and $m_3^2=0$ in (\ref{eq:f27}), respectively, and the integral
$Q^{3m}$ given by (\ref{eq:f25}).
The values 
of all the other integrals
appearing in Eqs. (\ref{eq:f28}) can be derived by taking appropriate zero$-$mass limits.

\subsection{Calculation of the integrals
               $P^{3m}$, $P^{2mh}$, and $P^{2me}$}
\label{sec:3}

Let us start by considering the integral $P^{3m}$ given by Eq. (\ref{eq:f27}).
Making the change $z\rightarrow 1-z$ in the second term on the
right$-$hand side in (\ref{eq:f27}), one finds that the first and second terms
are related to each other by the
$t\rightarrow s$ and
$m_2^2 \rightarrow m_4^2 $ interchanges.
Therefore, the expression for $P^{3m}$ takes the form
\begin{equation}
P^{3m}=\frac{1}{2}\,\left[ R(t,m_2^2,m_3^2)+R(s,m_4^2,m_3^2)\right] ~,\label{eq:f45}
\end{equation}
where
\begin{eqnarray}
\lefteqn{R(\alpha ,\beta ,m_3^2)=}\nonumber \\
& &{}=\frac{1}{\varepsilon_{IR}}
\int_{0}^{1}{\rm d}z\:(\beta -\alpha)
(-z\,\beta-(1-z)\,\alpha-{\rm i}\epsilon)^{\varepsilon_{IR}-1}
\nonumber \\
& &
\times {}_{2}F_{1}\left (1-\varepsilon_{IR}, \varepsilon_{IR};
1+\varepsilon_{IR};1-\frac{z(1-z)\, m_{3}^{2}+{\rm i}\epsilon}
{ z\,\beta+(1-z)\,\alpha +{\rm i}\epsilon}\right
).\nonumber \\ & &\label{eq:f46}
\end{eqnarray}
In order to evaluate this integral, we first apply the linear transformation
formula for the
hypergeometric functions:
\begin{eqnarray}
_2F_1(a,b;c;z)&=&\frac{\Gamma(c)\,\Gamma(c-a-b)}{\Gamma(c-a)\,\Gamma(c-b)}
\nonumber \\ & &\times
{}_2F_1(a,b;1+a+b-c;1-z)\nonumber \\
& & {}+
\frac{\Gamma(c)\,\Gamma(a+b-c)}{\Gamma(a)\,\Gamma(b)}(1-z)^{c-a-b}
\nonumber \\
& &\times {}_2F_1(c-a,c-b;1+c-a-b;1-z)
\nonumber \\
& & 
c-a-b\neq \pm n,\; \mid \mbox{arg}(1-z)\mid <\pi .\label{eq:f47}
\end{eqnarray}
With Eq. (\ref{eq:f47}) taken into account, (\ref{eq:f46}) becomes
\begin{eqnarray}
\lefteqn{R(\alpha,\beta,m_3^2)=}\nonumber \\
& &{}=\frac{1}{\varepsilon_{IR}}
\int_{0}^{1}{\rm d}z\:(\beta-\alpha)
(-z\,\beta-(1-z)\,\alpha-{\rm
i}\epsilon)^{\varepsilon_{IR}-1}\nonumber \\
& &\times \left[
\frac{\Gamma(1+\varepsilon_{IR})\,\Gamma(\varepsilon_{IR})}
{\Gamma(2\varepsilon_{IR})}\right.
\nonumber \\
& &\times {}_{2}F_{1}\left.\left (1-\varepsilon_{IR}, \varepsilon_{IR};
 1-\varepsilon_{IR}; \frac{z(1-z)\, m_{3}^{2}+{\rm i}\epsilon}
{ z\, \beta+(1-z)\, \alpha+{\rm i}\epsilon}\right ) \right.
\nonumber \\
& &{}-\left ( 
\frac{-z(1-z)\, m_{3}^{2}-{\rm i}\epsilon}
{ -z\, \beta-(1-z)\, \alpha-{\rm i}\epsilon}
\right ) ^{\varepsilon_{IR}}\nonumber \\
& &\times{}_{2}F_{1}\left. \left (2\varepsilon_{IR},1; 1+\varepsilon_{IR};
\frac{z(1-z)\, m_{3}^{2}+{\rm i}\epsilon}
{ z\, \beta+(1-z)\, \alpha+{\rm i}\epsilon}\right )
\right ].\label{eq:f48}
\end{eqnarray}
Assuming, with no loss of generality, that
\begin{equation}
\mid m_3^2 \mid  <  \mid \alpha \mid ,\, \mid \beta \mid 
\qquad \alpha /\beta > 0,\label{eq:f49}
\end{equation}
from which it follows 
\begin{equation}
\left |\frac{z(1-z)\, m_{3}^{2}+{\rm i}\epsilon}
{ z\,\beta +(1-z)\,\alpha+{\rm i}\epsilon}\right| <1 \qquad z\in [0,1],
\label{eq:f50}
\end{equation}
and making use of the series representation of the hypergeometric
function
\begin{equation}
_{2}F_{1}(a,b;c;z)=\sum_{n=0}^{\infty}\frac{\Gamma(a+n)\,\Gamma(b+n)\,
\Gamma(c)}{\Gamma(a)\,\Gamma(b)\,
\Gamma(c+n)}\frac{z^n}{n!},\;  \mid z\mid < 1,
\label{eq:f51}
\end{equation}
leads to the result
\begin{eqnarray}
\lefteqn{R(\alpha,\beta,m_3^2)=
\frac{\Gamma(\varepsilon_{IR})}{\Gamma(2\varepsilon_{IR})}
\left(1-\frac{\beta+{\rm i}\epsilon}{\alpha+{\rm i}\epsilon}\right)
\sum_{n=0}^{\infty}
\left( \frac{m_3^2+{\rm i}\epsilon}{\alpha+{\rm
i}\epsilon}\right)^n}
\nonumber \\
& &
\times \Bigg{\{ }(-\alpha-{\rm i}\epsilon)^{\varepsilon_{IR}}
\frac{\Gamma(n+\varepsilon_{IR})}{\Gamma(1+n)}\nonumber \\
& &
\times\!\!\int_0^1\!\!{\rm d}z\,\left[
z (1-z)
\right]^n
\left [
1-z\,\left ( 1-\frac{\beta+{\rm i}\epsilon}{\alpha+{\rm i}\epsilon}\right )
\right ]^{-(1+n-\varepsilon_{IR})}\nonumber \\
& &
-(-m_3^2-{\rm i}\epsilon)^{\varepsilon_{IR}}
\frac{\Gamma(n+2\varepsilon_{IR})}{\Gamma(1+n+\varepsilon_{IR})}\nonumber \\
& &
\times\!\!
\int_0^1\!\!{\rm d}z\,\left[ 
z (1-z)
\right]^{n+\varepsilon_{IR}}
\left [
1-z\,\left ( 1-\frac{\beta+{\rm i}\epsilon}{\alpha+{\rm i}\epsilon}\right )
\right ]^{-(1+n)}\Bigg{\} }.\nonumber \\ & &\label{eq:f52}
\end{eqnarray}
Taking into account the integral representation of the 
hypergeometric function given by Eq. (\ref{eq:f21}) yields
\begin{eqnarray}
\lefteqn{R(\alpha,\beta,m_3^2)=\frac{\Gamma(\varepsilon_{IR})}
{\Gamma(2\varepsilon_{IR})}
\left(1-\frac{\beta+{\rm i}\epsilon}{\alpha+{\rm i}\epsilon}\right)
\sum_{n=0}^{\infty}\left(\frac{m_3^2+{\rm
i}\epsilon}{\alpha+{\rm i}\epsilon}\right)^n}\nonumber \\
& & \times
\Bigg[(-\alpha-{\rm i}\epsilon)^{\varepsilon_{IR}}
\frac{\Gamma(n+\varepsilon_{IR})\,\Gamma(1+n)}{\Gamma(2+2n)}\nonumber \\
& & \times{}_2F_1\left (1+n-\varepsilon_{IR},1+n;2+2n;
1-\frac{\beta+{\rm i}\epsilon}{\alpha+{\rm i}\epsilon}\right ) \nonumber \\
& & {}-(-m_3^2-{\rm i}\epsilon)^{\varepsilon_{IR}}
\frac{\Gamma(n+2\varepsilon_{IR})\,\Gamma(1+n+\varepsilon_{IR})}
{\Gamma(2+2n+2\varepsilon_{IR})}\nonumber \\
& &\times
 {}_2F_1\left ( 1+n,1+n+\varepsilon_{IR};2+2n+2\varepsilon_{IR};1-
\frac{\beta+{\rm i}\epsilon}{\alpha+{\rm i}\epsilon}\right
)\Bigg].\nonumber \\
& &\label{eq:f53}
\end{eqnarray}
Although obtained under the restriction expressed by Eq. (\ref{eq:f49}), 
the above expression for $R(\alpha,\beta,m_3^2)$ can be
analytically continued for an arbitrary value of
$\beta$.
The series in (\ref{eq:f53}) is
convergent because it converges for
$\beta=0$ (see below) and the hypergeometric functions reach the
maximum at the same point.
To retain the convergence of the series, we still assume that $\mid m_3^2 
\mid < \mbox{$\mid \alpha\mid $}$.

By inspecting the sum on the right$-$hand side of (\ref{eq:f53}), we find that, 
if $\beta \neq 0$, all terms with the
exception of the first one ($n=0$) are of higher order in $\varepsilon_{IR}$
and can therefore
be omitted.
Upon replacing the whole sum by the first term,
and performing a few simple rearrangements, we find
\begin{eqnarray}
\lefteqn{R(\alpha,\beta \neq 0,m_3^2)=
\frac{\Gamma^2(\varepsilon_{IR})}{\Gamma(
2\varepsilon_{IR})}\, 
\left( 1-\frac{\beta+{\rm i}\epsilon}{\alpha+{\rm i}\epsilon}\right)
\Bigg[ (-\alpha-{\rm i}\epsilon)^{\varepsilon_{IR}}}\nonumber \\
& &\times
{}_2F_1\left (1-\varepsilon_{IR},1;2;
1-\frac{\beta+{\rm i}\epsilon}{\alpha+{\rm i}\epsilon}\right )\nonumber \\
& &{}-(-m_3^2-{\rm i}\epsilon)^{\varepsilon_{IR}}
\frac{1}{2\,(1+2\varepsilon_{IR})}\nonumber \\
& &\times{}_2F_1
\left (
1,1+\varepsilon_{IR};2+2\varepsilon_{IR};
1-\frac{\beta+{\rm i}\epsilon}{\alpha+{\rm i}\epsilon}
\right )\, \Bigg] +
\mathcal{O}
({\varepsilon_{IR}}).\nonumber \\ & & \label{eq:f54}
\end{eqnarray}
On the basis of formula (\ref{eq:f21}), one can easily show that the
first hypergeometric function on the right$-$hand side of the above
expression can be written in the form
\begin{eqnarray}
\lefteqn{_2F_1\left (1-\varepsilon_{IR},1;2;1-\frac{\beta+{\rm i}\epsilon}
{\alpha+{\rm i}\epsilon} \right )
=}\label{eq:f55}\\ & &{}\!\!\!=\frac{1}{\varepsilon_{IR}}\left (
1-\frac{\beta+{\rm i}\epsilon}{\alpha+{\rm
i}\epsilon}\right )^{-1}
\frac{(-\alpha-{\rm i}\epsilon)^{\varepsilon_{IR}}
-(-\beta-{\rm i}\epsilon)^{\varepsilon_{IR}}}
{(-\alpha-{\rm i}\epsilon)^{\varepsilon_{IR}}}.
\nonumber 
\end{eqnarray}
Now, inserting (\ref{eq:f55}) into (\ref{eq:f54}) leads to the expression
\begin{eqnarray}
\lefteqn{R(\alpha,\beta,m_3^2)=\frac{2 \Gamma^2(\varepsilon_{IR})}{\Gamma(
1+2\varepsilon_{IR})}
 \Bigg[ (-\alpha-{\rm i}\epsilon)^{\varepsilon_{IR}}\!\!-
(-\beta-{\rm i}\epsilon)^{\varepsilon_{IR}}}
\nonumber \\
& &{}-\frac{(-m_3^2-{\rm i}\epsilon)^{\varepsilon_{IR}}}{2}
\frac{\varepsilon_{IR}}
{1+2\varepsilon_{IR}} 
\left (1-\frac{\beta+{\rm i}\epsilon}{\alpha+{\rm i}\epsilon}\right )
\nonumber \\
& &\times{}_2F_1\left ( 1,1+\varepsilon_{IR};2+2\varepsilon_{IR};1-
\frac{\beta+{\rm i}\epsilon}{\alpha+{\rm i}\epsilon}\right ) \Bigg]
+\mathcal{O}
(\varepsilon_{IR}),\nonumber \\ & &
\label{eq:f56}
\end{eqnarray}
which is valid for arbitrary values of $\alpha$ and
$m_3^2$, and for $\beta \neq 0$.

On the other hand, if $\beta=0$, the first and second
hypergeometric functions appearing on the right$-$hand side of
Eq. (\ref{eq:f53}) reduce to
\[
\frac{\Gamma(2+2n)\,\Gamma(\varepsilon_{IR})}
{\Gamma(1+n)\,\Gamma(1+n+\varepsilon_{IR})}
\]
and
\[
\frac{\Gamma(2+2n+2\varepsilon_{IR})\,\Gamma(\varepsilon_{IR})}{\Gamma(1+n
+2\varepsilon_{IR})\,\Gamma(1+n+\varepsilon_{IR})},
\] 
respectively. This implies that when  $\beta =0$,
all terms in the sum are individually divergent and of the same order in
$\varepsilon_{IR}$ and, as such, all have to be taken into account, i. e.,
the summation has to be performed explicitly.
Therefore, as far as the expression for
$R(\alpha,\beta,m_3^2)$
given by Eq. (\ref{eq:f53}) is concerned,
the cases $\beta=0$ and $\beta\neq 0$ ought to be considered separately.

By setting $\beta=0$, Eq. (\ref{eq:f53}) takes the form
\begin{eqnarray}
\lefteqn{R(\alpha ,0,m_3^2)=}\nonumber \\
& &{}=\frac{\Gamma^2(\varepsilon_{IR})}
{\Gamma(
2\varepsilon_{IR})}\Bigg[\,(-\alpha-{\rm i}\epsilon)^{\varepsilon_{IR}}
\sum_{n=0}^{\infty}
\frac{1}{n+\varepsilon_{IR}}
\left ( \frac{m_3^2+{\rm i}\epsilon}{\alpha+{\rm i}\epsilon}
\right )^n
\nonumber \\
& &
{}-(-m_3^2-{\rm i}\epsilon)^{\varepsilon_{IR}}\sum_{n=0}^{\infty}
\frac{1}{n+2\varepsilon_{IR}}
\left ( \frac{m_3^2+{\rm i}\epsilon}
{\alpha+{\rm i}\epsilon}
\right )^n\,\Bigg]~.\label{eq:f57}
\end{eqnarray}
For the purpose of performing the summations in (\ref{eq:f57}), we note
that the series representation of the hypergeometric function given by
(\ref{eq:f51}) particularized for $a=1,\,c=1+b$ leads to the formula
\begin{equation}
_2F_1(1,b;1+b;z)=\sum_{n=0}^{\infty}\frac{b}{n+b}z^n
 \qquad \mid z \mid <1,\label{eq:f58}
\end{equation}
which, when taken into account in (\ref{eq:f57}), leads to the result
\begin{eqnarray}
\lefteqn{R(\alpha ,0,m_3^2)=\frac{2 \Gamma^2(\varepsilon_{IR})}{\Gamma(
1+2\varepsilon_{IR})}}\nonumber \\
& &{}\times
\left [\,(-\alpha -{\rm i}\epsilon)^{\varepsilon_{IR}}\,_2F_1\left (
1,\varepsilon_{IR};1+\varepsilon_{IR};
\frac{m_3^2+{\rm i}\epsilon}{\alpha+{\rm i}\epsilon}
\right ) 
\right.\nonumber \\
& &{}-
\left.
\frac{1}{2}(-m_3^2-{\rm i}\epsilon)^{\varepsilon_{IR}}\,_2F_1\left (
1,2\varepsilon_{IR};1+2\varepsilon_{IR};\frac{m_3^2+{\rm i}\epsilon}
{\alpha+{\rm i}\epsilon}
\right )\,\right ].\nonumber \\
& &\label{eq:f59}
\end{eqnarray}
This expression for $R(\alpha ,0,m_3^2)$ is valid for arbitrary values
of $\alpha$ and $m_3^2$.

To establish contact with the results obtained in Ref. \cite{bern},
instead of expanding everything in the expressions
(\ref{eq:f56}) and (\ref{eq:f59}) to the required order in 
$\varepsilon_{IR}$, let us, for the time being, expand 
only the occurring hypergeometric functions.
The relevant expansions are
\begin{eqnarray}
\lefteqn{z\,_2F_1(1,1+\delta;2+2\delta;z)=}
\nonumber \\ & &{}=\frac{1+2\delta}{\delta}
\Big[1-(1-z)^{\delta}-2\,\delta ^2\,\mbox{Li}_2(z)+\mathcal{O}(\delta
^3)\Big] \label{eq:f60}
\end{eqnarray}
and
\begin{equation}
_2F_1(1,\delta;1+\delta;z)=
1-\delta~ \ln(1-z)-\delta ^2\,\mbox{Li}_2(z)+\mathcal{O}(\delta
^3)~,\label{eq:f61}
\end{equation}
where $\mbox{Li}_2 (z)$ stands for the Euler dilogarithm \cite{hill} defined
as
\begin{equation}
{\rm {Li}}_2(z)=-\int_0^1 \frac{dt}{t}\;\ln(1-tz)~.\label{eq:f41}
\end{equation}
Taking (\ref{eq:f60}) into account in (\ref{eq:f56}), we find that
\begin{eqnarray}
\lefteqn{R(\alpha,\beta ,m_3^2)=\frac{2 \Gamma^2(\varepsilon_{IR})}
{\Gamma(1+2\varepsilon_{IR})}
 \Bigg[ (-\alpha-{\rm i}\epsilon)^{\varepsilon_{IR}}\!\!
-(-\beta-{\rm i}\epsilon)^{\varepsilon_{IR}}}
\nonumber \\
& &
{}-\frac{1}{2}
(-m_3^2-{\rm i}\epsilon)^{\varepsilon_{IR}}+\frac{1}{2}
\frac{(-\beta-{\rm i}\epsilon)^{\varepsilon_{IR}}
(-m_3^2-{\rm i}\epsilon)^{\varepsilon_{IR}}}
{(-\alpha-{\rm i}\epsilon)^{\varepsilon_{IR}}}\nonumber \\
& &
{}+\varepsilon_{IR}^2\mbox{Li}_2\left (
1-\frac{\beta+{\rm i}\epsilon}{\alpha+{\rm i}\epsilon}\right )\Bigg]
+{\mathcal{O}}(\varepsilon_{IR}).\label{eq:f62}
\end{eqnarray}
Next, using the expansion (\ref{eq:f61}), the expression (\ref{eq:f59}) 
becomes
\begin{eqnarray}
\lefteqn{R(\alpha,0,m_3^2)=}\nonumber \\
& &{}=\frac{2 \Gamma^2(\varepsilon_{IR})}
{\Gamma(1+2\varepsilon_{IR})}
\Bigg\{ (-\alpha-{\rm i}\epsilon)^{\varepsilon_{IR}}-\frac{1}{2}
(-m_3^2-{\rm i}\epsilon)^{\varepsilon_{IR}}
\nonumber \\
& &
{}+\varepsilon_{IR}^2
\left [ -\mbox{Li}_2
\left ( 1-\frac{m_3^2+{\rm i}\epsilon}
{\alpha+{\rm i}\epsilon}
\right )
+\frac{\pi^2}{6}\right]
\Bigg\} +\mathcal{O}(\varepsilon_{IR})~.
\label{eq:f63}
\end{eqnarray}
In arriving at (\ref{eq:f63}), the relation
\begin{equation}
\mbox{Li}_2(1-z)=
-\mbox{Li}_2(z)-\ln(z)\,\ln(1-z)+\frac{\pi^2}{6}~,\label{eq:f64}
\end{equation}
has been employed.
On the basis of Eqs. (\ref{eq:f62}) and (\ref{eq:f45}) 
we find that the integral $P^{3m}$
is given by
\begin{eqnarray}
\lefteqn{P^{3m}=
\frac{\Gamma^2(\varepsilon_{IR})}{\Gamma(1+2\varepsilon_{IR})}
\Bigg\{ \,(-s-{\rm i}\epsilon)^{\varepsilon_{IR}}
+(-t-{\rm i}\epsilon)^{\varepsilon_{IR}}}\nonumber \\
& & {}-(-m_2^2-{\rm i}\epsilon)^{\varepsilon_{IR}}
-(-m_3^2-{\rm i}\epsilon)^{\varepsilon_{IR}}
-
(-m_4^2-{\rm i}\epsilon)^{\varepsilon_{IR}}\nonumber \\
& & {}+\frac{1}{2}
\frac{(-m_2^2-{\rm i}\epsilon)^{\varepsilon_{IR}}
(-m_3^2-{\rm i}\epsilon)^{\varepsilon_{IR}}}
{(-t-{\rm i}\epsilon)^{\varepsilon_{IR}}}\nonumber \\& & {}+
\frac{1}{2} \frac{(-m_3^2-{\rm i}\epsilon)^{\varepsilon_{IR}}
(-m_4^2-{\rm i}\epsilon)^{\varepsilon_{IR}}}{
(-s-{\rm i}\epsilon)^{\varepsilon_{IR}}}
\nonumber \\
& &
{}+\varepsilon_{IR}^{2}\left[ 
\mbox{Li}_2\left (
1-\frac{m_2^2+{\rm i}\epsilon}{t+{\rm i}\epsilon}\right )+ 
\mbox{Li}_2\left (
1-\frac{m_4^2+{\rm i}\epsilon}{s+{\rm i}\epsilon}\right )\right]\,\Bigg\} 
\nonumber \\& &{}+
{\mathcal{O}}(\varepsilon_{IR}).\label{eq:f65}
\end{eqnarray}
Similarly, combining Eqs. (\ref{eq:f62}), (\ref{eq:f63}), and (\ref{eq:f45}), 
we obtain
\begin{eqnarray}
P^{2mh} &=& \frac{\Gamma^2(\varepsilon_{IR})}{\Gamma(1+2\varepsilon_{IR})}
\Bigg\{ \,(-s-{\rm i}\epsilon)^{\varepsilon_{IR}}
+(-t-{\rm i}\epsilon)^{\varepsilon_{IR}}\nonumber \\
& &{}-(-m_3^2-{\rm i}\epsilon)^{\varepsilon_{IR}}
-
(-m_4^2-{\rm i}\epsilon)^{\varepsilon_{IR}}\nonumber \\
& & {}+\frac{1}{2} \frac{(-m_3^2-{\rm i}\epsilon)^{\varepsilon_{IR}}
(-m_4^2-{\rm i}\epsilon)^{\varepsilon_{IR}}}
{(-s-{\rm i}\epsilon)^{\varepsilon_{IR}}}\nonumber \\
& & {}+\varepsilon_{IR}^{2}\left[ \frac{\pi^2}{6}
- 
\mbox{Li}_2\left (
1-\frac{m_3^2+{\rm i}\epsilon}{t+{\rm i}\epsilon}\right )\right.\nonumber \\
& & {}+\mbox{Li}_2\left.\left (
1-\frac{m_4^2+{\rm i}\epsilon}{s+{\rm i}\epsilon}\right )\right]\,\Bigg\}
+\mathcal{O}(\varepsilon_{IR}). \label{eq:f66}
\end{eqnarray}
We now turn to evaluate the integral $P^{2me}$.
To this end, we set $m_3^2=0$ in Eq. (\ref{eq:f27}). 
As a result, both
hypergeometric functions appearing in (\ref{eq:f27}) reduce to
\begin{equation}
_2F_1(1-\varepsilon_{IR},\varepsilon_{IR};1+\varepsilon_{IR};1)=
\frac{\Gamma(1+\varepsilon_{IR})\,\Gamma(\varepsilon_{IR})}
{\Gamma(2\varepsilon_{IR})},\label{eq:f67}
\end{equation}
making it possible to perform the remaining integration analytically.
The exact result for the integral $P^{2me}$ is
\begin{eqnarray}
P^{2me}&=&
\frac{\Gamma^2(\varepsilon_{IR})}{\Gamma(1+2\varepsilon_{IR})}
 \Big[\,(-s-{\rm i}\epsilon)^{\varepsilon_{IR}}
+(-t-{\rm i}\epsilon)^{\varepsilon_{IR}}\nonumber \\
& &{}-(-m_2^2-{\rm i}\epsilon)^{\varepsilon_{IR}}
-(-m_4^2-{\rm i}\epsilon)^{\varepsilon_{IR}}   \,\Big].
\label{eq:f68}
\end{eqnarray}
It is valid for arbitrary values of $m_2^2$ and $m_4^2$. Consequently,
the expressions for the integrals $P^{1m}$ and $P^{0m}$ are obtained by
setting $m_2^2=0$ and $m_2^2=m_4^2=0$, respectively, in (\ref{eq:f68}).

A remark concerning the issue of the zero$-$mass limits of the massive
integrals $P^K$ is in order. By looking at the expressions for the integrals 
$P^K$ given above, one observes that the limits $P^{2mh}\rightarrow P^{1m}$,
$P^{3m}\rightarrow P^{2me}$, and $P^{3m}\rightarrow P^{2mh}$ are not smooth.
On the other hand, the limits $P^{2me}\rightarrow P^{1m}$ and 
$P^{1m}\rightarrow P^{0m}$ are smooth. In general, there is no reason for the
zero$-$ mass limits to be smooth. Namely, the limit of taking a mass to zero
does not necessarily commute with the $1/{\varepsilon}_{IR}$ expansion of the
dimensional regularization, which has been truncated at 
$\mathcal{O}({\varepsilon}_{IR}^0)$. Note, however, that if we were able to evaluate the integral
$P^{3m}$ in (\ref{eq:f27}) analytically for general ${\varepsilon}_{IR}$, the
result thus obtained would suffice to obtain the results for the other integrals
$P^{2mh}, P^{2me}, P^{1m}$, and $P^{0m}$ by simply setting $m_2^2=0$, $m_3^2=0$,
 $m_2^2=m_3^2=0$, and $m_2^2=m_3^2=m_4^2=0$,
respectively, and then expanding these results to the required order in 
${\varepsilon}_{IR}$.

\subsection{Calculation of the integral $Q^{3m}$}
\label{sec:4}

As it is seen from (\ref{eq:f25}), the integral $Q^{3m}$
is given in terms of two hypergeometric functions, both of which can be
conveniently written in the form
\begin{eqnarray}
& & {}_{2}F_{1}
\left (1-\varepsilon_{IR}, \varepsilon_{IR}; 1+\varepsilon_{IR};
 1-\frac{z(1-z)\, m_{3}^{2}+{\rm i}\epsilon}
{ z\, \alpha +(1-z)\, \beta +{\rm i}\epsilon}
\right )\nonumber \\
& &\qquad \qquad \qquad
(\alpha ,\beta )\in \Big{\{ } (m_2^2,t),(s,m_4^2)\Big{\}}.\label{eq:f30}
\end{eqnarray} 
Making use of the transformation formula
\begin{equation}
\,_{2}F_{1}(a,b;c;z)=(1-z)^{c-a-b}\,_{2}F_{1}(c-a,c-b;c;z),\label{eq:f31}
\end{equation}
the symmetry of $_{2}F_{1}$ with respect to the arguments $a$ and $b$, i.e.,
\begin{equation}
\,_{2}F_{1}(a,b;c;z)\,=\,_{2}F_{1}(b,a;c;z),\label{eq:f32}
\end{equation}
and the integral representation
of the hypergeometric function (\ref{eq:f21}), we can write
the hypergeometric function given by (\ref{eq:f30}) in the form
\begin{eqnarray}
& &{}_{2}F_{1}\left (1-\varepsilon_{IR}, \varepsilon_{IR}; 1+\varepsilon_{IR};
 1-\frac{z(1-z)\, m_{3}^{2}+{\rm i}\epsilon}
{ z\, \alpha +(1-z)\, \beta +{\rm i}\epsilon}\right )=\nonumber \\
& &=\left (\frac{-z(1-z)\, m_{3}^{2}-{\rm i}\epsilon}{-z\, 
\alpha -(1-z)\, \beta
-{\rm i}\epsilon}\right )^{\varepsilon_{IR}}
\frac{\Gamma(1+\varepsilon_{IR})}{\Gamma(1-\varepsilon_{IR})\,
\Gamma(2 \varepsilon_{IR})}
\nonumber \\
& &\times  
\int_{0}^{1}{\rm d}y\, 
y^{2 \varepsilon_{IR}-1}(1-y)^{-\varepsilon_{IR}}
\nonumber \\ & &{}\times\left [1-y\left ( 1-
\frac{z(1-z)\, m_{3}^{2}+{\rm i}\epsilon}{z\, \alpha +(1-z)\, \beta
+{\rm i}\epsilon}\right )\right ]^{-1}.\label{eq:f33}
\end{eqnarray}
Upon substituting (\ref{eq:f33}) into (\ref{eq:f25}), 
and utilizing the identity (\ref{eq:f23}) 
once more, we find that
\begin{eqnarray}
\lefteqn{Q^{3m}=
\frac{\Gamma(\varepsilon_{IR})}{2\,\Gamma(1-\varepsilon_{IR})\,
\Gamma(2 \varepsilon_{IR})}
\int_{0}^{1}{\rm d}z
\frac{1}{z-z_0}}\nonumber \\
& &{}\times[-z(1-z)\, m_{3}^{2}-{\rm
i}\epsilon]^{\varepsilon_{IR}}
\int_0^1{\rm d}y
\,y^{2
\varepsilon_{IR}}(1-y)^{-\varepsilon_{IR}}\nonumber \\
& &
{}\times
\left\{
\left[-z(1-z)\,
m_{3}^{2}+z\,s+(1-z)\,m_4^2\right]\right.\nonumber
\\ & &{}\times
\left[-z\,s-(1-z)\,m_4^2-{\rm i}\epsilon \right.\nonumber
\\ & &
{}+y\,\left.(-z(1-z)\,
m_{3}^{2}+z\,s+(1-z)\,m_4^2)\right]^{-1}\nonumber \\
& &
{}-
\left[-z(1-z)\,
m_{3}^{2}+z\,m_2^2+(1-z)\,t\right]\nonumber \\ & &
{}\times\left[-z\,m_2^2-(1-z)\,t-{\rm i}\epsilon
\right.\nonumber 
\\& &{}+y\left.\left.(-z(1-z)\,
m_{3}^{2}+z\,m_2^2+(1-z)\,t)\right]^{-1}
\right\}~.\label{eq:f34}
\end{eqnarray}
Since we are interested in obtaining the value of $Q^{3m}$ to
$\mathcal{O}(\varepsilon_{IR}^0)$, the fact that the expansion
of the prefactor in the above expression is of the form $1+\mathcal{O}(\varepsilon_{IR})$,
allows us to set $\varepsilon_{IR}=0$ in the integrand
in (\ref{eq:f34}). As a result, the 
expression for $Q^{3m}$ reduces to
\begin{eqnarray}
\lefteqn{Q^{3m} = \int_{0}^{1}{\rm d}z
\frac{1}{z-z_0}\int_0^1{\rm d}y}
\label{eq:f36}\\
& &
\times\left\{
\left[-z(1-z)\,
m_{3}^{2}+z\,s+(1-z)\,m_4^2\right]\right.\nonumber
\\& &\times\left[-z\,s-(1-z)\,m_4^2-{\rm i}\epsilon
\right.\nonumber \\
& &{}+y\,\left.(-z(1-z)\, m_{3}^{2}+z\,s+(1-z)\,m_4^2)
\right]^{-1}
\nonumber \\
& &
{}-
\left[-z(1-z)\,
m_{3}^{2}+z\,m_2^2+(1-z)\,t\right]\nonumber \\
& &\times\left[-z\,m_2^2-(1-z)\,t-{\rm i}\epsilon
\right.\nonumber \\& &{}+y\left.\left.(-z(1-z)\,
m_{3}^{2}+z\,m_2^2+(1-z)\,t)\right]^{-1}
\right\}\!\!+\mathcal{O}(\varepsilon_{IR}).\nonumber 
\end{eqnarray}
Performing the $y$ integration, we find
\begin{equation}
Q^{3m}= \int_0^1\frac{{\rm d}z}{z-z_0}
\ln
\left (
\frac
{z(t-m_2^2)-t-{\rm i}\epsilon}
{z(m_4^2-s)-m_4^2-{\rm i}\epsilon}
\right )+\mathcal{O}(\varepsilon_{IR})~\cdot \label{eq:f38}
\end{equation}
To carry out the remaining integration, it is important to note
that the residue of the integrand at the pole $z_0$ is zero,
and the logarithm does not cross the cut.
This fact allows us to make a few 
simple transformations of the integrand.

Thus, upon the substitution $z\rightarrow z+z_0$,
the decomposition
\[
\int_0^1dz \rightarrow -\int_0^{\displaystyle {-z_0}}dz+
\int_0^{\displaystyle {1-z_0}}dz~,
\]
followed by a change of the variable $z\rightarrow -z\,z_0$ in the first,
and $z\rightarrow z\,(1-z_0)$ in the second term, the integral $Q^{3m}$
can be written down as
\begin{eqnarray}
Q^{3m}&=&
\int_0^1 \frac{{\rm d}z}{z}\Big{\{ }\,\ln\big[ 1-z\,\big( 1-(m_2^2
+{\rm i}\epsilon)f^{3m}\big)\big]\nonumber \\
& &+\ln\big[ 1-z\,\big( 1-(m_4^2
+{\rm i}\epsilon)f^{3m}\big)\big]-\nonumber \\
& &-\ln\big[ 1-z\,\big( 1-(s
+{\rm i}\epsilon)f^{3m}\big)\big]\nonumber \\
& &-\ln\big[ 1-z\,\big( 1-(t
+{\rm i}\epsilon)f^{3m}\big)\big]\,\Big{\} }+\mathcal{O}(\varepsilon_{IR}),
\label{eq:f39}
\end{eqnarray}
where we have introduced the abbreviation
\begin{equation}
f^{3m}=\frac{s+t-m_2^2-m_4^2}{st-m_2^2m_4^2}~.\label{eq:f40}
\end{equation}
Expressed in terms of the Euler dilogarithm, 
the final result for the integral $Q^{3m}$  
takes the form
\begin{eqnarray}
\lefteqn{Q^{3m}= 
\mbox{Li}_2\Big[1-(s+{\rm i}\epsilon)f^{3m}\Big]\!
+\!\mbox{Li}_2\Big[1-(t+{\rm i}\epsilon)f^{3m}\Big] }\label{eq:f42} \\
& &
\!\!\!\!\!\!{}-\!\mbox{Li}_2\Big[1-(m_2^2+{\rm i}\epsilon)f^{3m}\Big]\!\!
-\!\mbox{Li}_2\Big[1-(m_4^2+{\rm i}\epsilon)f^{3m}\Big]
\!\!+\!\mathcal{O}(\varepsilon_{IR}).\nonumber
\end{eqnarray}
The expression for $Q^{3m}$ does not depend on $m_3^2$,
and is valid for arbitrary values of
$m_2^2$ and $m_4^2$. A consequence of this is that the expression for
the integral $Q^{2mh}$ to the same order in ${\varepsilon}_{IR}$
can simply be obtained by setting $m_2^2=0$ in (\ref{eq:f42}). 
Note, however, that,
strictly speaking, the expressions for $Q^{2me}$, $Q^{1m}$, and $Q^{0m}$
cannot be obtained by taking appropriate zero$-$mass limits of the same
expression. Namely, the expression (\ref{eq:f34}) has been derived assuming
that $m_3^2\neq0$. Therefore, the
${\varepsilon}_{IR}$ expansion of (\ref{eq:f34}) 
is not justified in the $m^2_3\rightarrow 0$ limit. In order to 
obtain the integrals $Q^{2me}$, $Q^{1m}$, and $Q^{0m}$, we proceed as
follows. We return to Eq. (\ref{eq:f25}) 
and set $m_3^2=0$. As a result, Eq. (\ref{eq:f25})
reduces to
\begin{eqnarray}
Q^{2me}&=&
\frac{\Gamma^2(\varepsilon_{IR})}
{2\,\Gamma(2 \varepsilon_{IR})}\!\int_{0}^{1}\!\!\!\!
{\rm d}z\:
\frac{1}{z-z_0}
[(-z\, m_2^2-(1-z)\, t-{\rm i}\epsilon)^{\varepsilon_{IR}}\nonumber \\
& &  
{}-(-z\,s-(1-z)\,m_4^2-{\rm i}\epsilon)^{\varepsilon_{IR}}].\label{eq:f43}
\end{eqnarray}
Expanding this expression into power series in 
${\varepsilon}_{IR}$
\begin{equation}
Q^{2me}=
\int_0^1\frac{{\rm d}z}{z-z_0}
\ln
\left (
\frac
{z(t-m_2^2)-t-{\rm i}\epsilon}
{z(m_4^2-s)-m_4^2-{\rm i}\epsilon}
\right )
\!\!+\mathcal{O}(\varepsilon_{IR}),\label{eq:f44}
\end{equation}
and comparing it with the expansion of $Q^{3m}$ given
by Eq. (\ref{eq:f38}), 
we find that the expansions for $Q^{3m}$ and $Q^{2me}$ coincide 
to the required order in
${\varepsilon}_{IR}$.
This being the case, all other intgerals $Q^K$ can be derived from
the integral $Q^{3m}$ given by Eq. (\ref{eq:f42}) 
by taking appropriate zero$-$mass limits.

\subsection{Results}
\label{sec:5}

Having obtained, in the preceding subsections, the closed form expressions
for the integrals $P^K(s,t;\{m_i^2\})$ and
\linebreak $Q^K(s,t; \{m_i^2\})$ to order
$\mathcal{O}$$(\varepsilon_{IR}^0)$, we are now in a position to write down
explicit expressions for the integrals
$I_4^K(s,t;\{m_i^2\})$.

Before proceeding, it is convenient to introduce the functions
\begin{eqnarray}
f^{3m}=f^{2me}&=&\frac{s+t-m_2^2-m_4^2}{st-m_2^2m_4^2}~,
\nonumber \\
f^{2mh}=f^{1m}&=&\frac{s+t-m_4^2}{st}~,\nonumber \\
f^{0m}&=&\frac{s+t}{st}~,\label{eq:f69}
\end{eqnarray}
and the notation
\begin{equation}
r_{\Gamma}=
\frac{\Gamma(1-\varepsilon_{IR}) {\Gamma}^2(1+\varepsilon_{IR})}
{\Gamma(1+2\varepsilon_{IR})} \label{eq:f70}
\end{equation}
for the $\Gamma$ function prefactor.

Now, on the basis of Eqs. (\ref{eq:f28}), (\ref{eq:f42}),
 (\ref{eq:f65}), (\ref{eq:f66}), and (\ref{eq:f68}),
we find that\\
the three$-$mass scalar box integral is
\begin{eqnarray}
\lefteqn{I^{3m}_{4}(s,t;m_2^2,m_3^2,m_4^2)=
\frac{{\rm i}}{(4\pi)^{2}}\,
\frac{r_{\Gamma}}{s\, t-m_{2}^{2}\, m_{4}^{2}} 
}\nonumber\\ & &\times\Bigg\{ \frac{2}{\varepsilon_{IR}^{2}}
\Bigg[
\left (\frac {-s-{\rm i}\epsilon}{4\pi {\mu}^2}\right )^{\varepsilon_{IR}}
\!\!\!\!+\!
\left (\frac {-t-{\rm i}\epsilon}{4\pi {\mu}^2}\right )^{\varepsilon_{IR}}
\!\!\!\!-\!\left (\frac {-m_2^2-{\rm i}\epsilon}{4\pi {\mu}^2}\right )^{\varepsilon_{IR}}
\nonumber \\ & &{}-\!
\left (\frac {-m_3^2-{\rm i}\epsilon}{4\pi {\mu}^2}\right )^{\varepsilon_{IR}}
\!\!\!\!-\!
\left (\frac {-m_4^2-i\epsilon}{4\pi {\mu}^2}\right )^{\varepsilon_{IR}}
\, \Bigg] \nonumber \\
& &
{}+\frac{1}{\varepsilon_{IR}^{2}}\!
\left (\frac {-m_2^2-{\rm i}\epsilon}{4\pi {\mu}^2}\right )^{\varepsilon_{IR}}
\!\!\left (\frac {-m_3^2-{\rm i}\epsilon}{4\pi {\mu}^2}\right )^{\varepsilon_{IR}}
\!\!\left (\frac {-t-{\rm i}\epsilon}{4\pi {\mu}^2}\right )^{-\varepsilon_{IR}}
\nonumber \\
& &{}+\frac{1}{\varepsilon_{IR}^{2}}\!
\left (\frac {-m_3^2-{\rm i}\epsilon}{4\pi {\mu}^2}\right )^{\varepsilon_{IR}}
\!\!\left (\frac {-m_4^2-{\rm i}\epsilon}{4\pi {\mu}^2}\right )^{\varepsilon_{IR}}
\!\!\left (\frac {-s-{\rm i}\epsilon}{4\pi {\mu}^2}\right )^{-\varepsilon_{IR}}
\nonumber \\
& &
{}+2\, \mbox{Li}_{2}\left(\, 1-\frac{m_{2}^{2}+{\rm i}\epsilon }{t+{\rm i}\epsilon}\,
\right)+2\,\mbox{Li}_{2}\left(\, 1-\frac{m_{4}^{2}+{\rm i}\epsilon}
{s+{\rm i}\epsilon}\, \right)
\nonumber \\
& &
{}+2\, \mbox{Li}_{2}\Big[\, 1-(s+{\rm i}\epsilon)\, f^{3m}\,\Big]
+2\, \mbox{Li}_{2}\Big[\,
1-(t+{\rm i}\epsilon)\, f^{3m}\, \Big]
\nonumber \\
& &
{}-2\, \mbox{Li}_{2}\Big[\, 1-(m_{2}^{2}+{\rm i}\epsilon)\, f^{3m}\, \Big]
\nonumber \\& &{}-2\,
\mbox{Li}_{2}\Big[\,
1-(m_{4}^{2}+{\rm i}\epsilon)\, f^{3m}\, 
\Big] \Bigg\}
+ \mathcal{O}({\epsilon}_{IR})~, \label{eq:f71}
\end{eqnarray}
the adjacent ("hard") two$-$mass scalar box integral is
\begin{eqnarray}
\lefteqn{I_4^{2mh}(s,t;m_3^2,m_4^2) =
\frac{{\rm i}}{(4\pi)^{2}}\,
\frac{r_{\Gamma}}{s\, t}
\,\Bigg\{
\, \frac{2}{\varepsilon_{IR}^{2}}\,
\Bigg [
\,
\left (\frac {-s-{\rm i}\epsilon}{4\pi {\mu}^2}\right )^{\varepsilon_{IR}}
}\nonumber \\ & & {}+\!
\left (\frac {-t-{\rm i}\epsilon}{4\pi {\mu}^2}\right )^{\varepsilon_{IR}}
\!\!\!\!-\!
\left (\frac {-m_3^2-{\rm i}\epsilon}{4\pi {\mu}^2}\right )^{\varepsilon_{IR}}
\!\!\!\!-\!
\left (\frac {-m_4^2-{\rm i}\epsilon}{4\pi {\mu}^2}\right )^{\varepsilon_{IR}}
\Bigg ]
\nonumber \\
& &
{}+\frac{1}{\varepsilon_{IR}^{2}}
\left (\frac {-m_3^2-{\rm i}\epsilon}{4\pi {\mu}^2}\right )^{\varepsilon_{IR}}
\!\!
\left (\frac {-m_4^2-{\rm i}\epsilon}{4\pi {\mu}^2}\right )^{\varepsilon_{IR}}
\!\!
\left (\frac {-s-{\rm i}\epsilon}{4\pi {\mu}^2}\right )^{-\varepsilon_{IR}}
\nonumber \\
& &
{}+2\,\mbox{Li}_{2}\left(\, 1-\frac{m_{4}^{2}+{\rm i}\epsilon }{s+{\rm i}\epsilon}\,
\right)-2\,\mbox{Li}_{2}\left(\, 1-\frac{m_{3}^{2}+{\rm i}\epsilon}
{t+{\rm i}\epsilon}\, \right)
\nonumber \\
& &
{}+2\, \mbox{Li}_{2}\Big[\, 1-(s+{\rm i}\epsilon)\, f^{2mh}\, \Big]
+2\, \mbox{Li}_{2}\Big[\,
1-(t+{\rm i}\epsilon)\, f^{2mh}\,  \Big]
\nonumber \\
& &
{}-2\, \mbox{Li}_{2}\Big[\, 1-(m_{4}^{2}+{\rm i}\epsilon)\, f^{2mh}\, \Big]
\, \Bigg\}
+ \mathcal{O}({\epsilon}_{IR})~,\label{eq:f72}
\end{eqnarray}
the opposite ("easy") two$-$mass scalar box integral is
\begin{eqnarray}
\lefteqn{I^{2me}_{4}(s,t;m_2^2,m_4^2) =
\frac{{\rm i}}{(4\pi)^{2}}\,
\frac{r_{\Gamma}}{s\, t-m_{2}^{2}\, m_{4}^{2}}\,
}\nonumber \\
& &{}\Bigg\{ \, \frac{2}{\varepsilon_{IR}^{2}}\,
\Bigg [
\left (\frac {-s-{\rm i}\epsilon}{4\pi {\mu}^2}\right )^{\varepsilon_{IR}}
\!\!\!+\!
\left (\frac {-t-{\rm i}\epsilon}{4\pi {\mu}^2}\right )^{\varepsilon_{IR}}
\!\!\!-\!
\left (\frac {-m_2^2-{\rm i}\epsilon}{4\pi {\mu}^2}\right )^{\varepsilon_{IR}}
\nonumber \\
& &{}-
\left (\frac {-m_4^2-{\rm i}\epsilon}{4\pi {\mu}^2}\right )^{\varepsilon_{IR}}
\,
\Bigg ]+2\, \mbox{Li}_{2}\Big[\, 1-(s+{\rm i}\epsilon)\, f^{2me}\, \Big]
 \nonumber \\
& &
{}
+2\, \mbox{Li}_{2}\Big[\,
1-(t+{\rm i}\epsilon)\, f^{2me}\, \Big]
-2\, \mbox{Li}_{2}\Big[\, 1-(m_{2}^{2}+{\rm i}\epsilon)\, f^{2me}\, \Big]
\nonumber \\
& &
{}
-2\, \mbox{Li}_{2}\Big[\,
1-(m_{4}^{2}+{\rm i}\epsilon)\, f^{2me}\, \Big]\, \Bigg\}
+ \mathcal{O}({\epsilon}_{IR})~.\label{eq:f73}
\end{eqnarray}
An essential feature of the above expression for 
$I^{2me}_{4}$
is that it is well behaved in the
$m_2^2 \rightarrow 0$,
$m_4^2 \rightarrow 0$ limits.
A consequence of this is that it contains both the one$-$mass scalar box
integral
and the massless scalar box integral.
Thus, setting $m_2^2=0$ in (\ref{eq:f73}) and making use of the fact that
$\rm {Li}_2(1)$=${\pi}^2/6$, we find that 
the one$-$mass box scalar integral  is
\begin{eqnarray}
\lefteqn{I^{1m}_{4}(s,t;m_4^2)=
\frac{{\rm i}}{(4\pi)^{2}}\,
\frac{r_{\Gamma}}{s\,t}\,\Bigg{\{} 
\, \frac{2}{\varepsilon_{IR}^{2}}\,
\left [\,
\left (\frac {-s-{\rm i}\epsilon}{4\pi {\mu}^2}\right )^{\varepsilon_{IR}}
\right. }\nonumber \\
& &{}+
\left.\left (\frac {-t-{\rm i}\epsilon}{4\pi {\mu}^2}\right )^{\varepsilon_{IR}}-
\left (\frac {-m_4^2-{\rm i}\epsilon}{4\pi {\mu}^2}\right )^{\varepsilon_{IR}}\,
\right ]
\nonumber \\
& &
{}+2\,\mbox{Li}_{2}\Big[\, 1-(s+{\rm i}\epsilon)\, f^{1m} \,\Big]
+2\,\mbox{Li}_{2}\Big[\,
1-(t+{\rm i}\epsilon)\,f^{1m} \, \Big]
\nonumber \\
& &{}-2\,\mbox{Li}_{2}\Big[\, 1-(m_{4}^{2}+{\rm i}\epsilon)\,f^{1m}\, \Big]
-\frac{\pi^{2}}{3}\,\Bigg\}
+ \mathcal{O}({\epsilon}_{IR}). \label{eq:f74}
\end{eqnarray}
Finally, setting $m_4^2 = 0$ in (\ref{eq:f74}), we find that
 the massless scalar box integral is given by
\begin{eqnarray}
\lefteqn{I^{0m}_{4}(s,t) =
\frac{{\rm i}}{(4\pi)^{2}}\,
\frac{r_{\Gamma}}{s\, t}\,\Bigg{\{}  
\, \frac{2}{\varepsilon_{IR}^{2}}\,
\left [
\left (\frac {-s-{\rm i}\epsilon}{4\pi {\mu}^2}\right )^{\varepsilon_{IR}}
\right.}\nonumber \\
& &{}+
\left.\left (\frac {-t-{\rm i}\epsilon}{4\pi {\mu}^2}\right )^{\varepsilon_{IR}}\,
\right ] +2\,\mbox{Li}_{2}\Big[\, 1-(s+{\rm i}\epsilon)\,f^{0m}\, \Big]
 \nonumber \\
& &{}
+2\,\mbox{Li}_{2}\Big[\,
1-(t+{\rm i}\epsilon)\,f^{0m} \, \Big]
-2\, \frac{\pi^{2}}{3}\,\Bigg\}
+ \mathcal{O}({\epsilon}_{IR}).
\label{eq:f75}
\end{eqnarray}
The above expressions for the one$-$loop IR divergent scalar box integrals 
$I_4^K(s,t;\{m_i^2\})$ constitute the main result of this paper.
It is important to emphasize that,
owing to the fact that we have kept the "causal"
${\rm i}\epsilon$ systematically throughout the calculation, these expressions
are valid for arbitrary values of the relevant kinematic
variables: external masses $m_i^2~(i=2,3,4)$ and the Mandelstam
variables $s$ and $t$.

\noindent
As stated in the Introduction, the integrals
$I_4^K(s,t,\{m_i^2\})$ have been evaluated in Ref. \cite{bern} with the help of the
partial differential equation technique. The calculation has been
performed in the Euclidean region, where all kinematic variables
are negative, i.e.,
\begin{equation}
s,~t<0,~~m_2^2,~m_3^2,~m_4^2<0~,\label{eq:f76}
\end{equation}
and the results thus obtained have been analytically continued to the
positive values of the kinematic variables (physical region) by applying
the following replacements:
\begin{equation}
s\rightarrow s+{\rm i}\epsilon~,~~
t\rightarrow t+{\rm i}\epsilon~,~~
m_i^2 \rightarrow m_i^2+{\rm i}\epsilon~.\label{eq:f77}
\end{equation}
In order to facilitate the
comparison of our results with those of Ref. \cite{bern}, we have written
our results in the same form in which they have been presented in Ref.
\cite{bern}.
A glance at the expressions (\ref{eq:f71}$-$\ref{eq:f75}) reveals that 
they all have the same general form, namely,
\begin{eqnarray}
I_{\displaystyle 4}^K
(s,t; \{m_i^2\})
&=&\frac{{\rm i}}{(4\pi )^2}\,
r_{\Gamma}
g^K
\left [
\frac
{G^K
(s,t;{\varepsilon}_{IR}; \{m_i^2\})}{{\varepsilon}_{IR}^2}
\right. \nonumber \\
& &{}+H^K
(s,t; \{m_i^2\})\left.
\right ]
+\mathcal{O}({\epsilon}_{IR})
\nonumber \\
& &
K\in \{3m,2mh,2me,1m,0m\}.\label{eq:f78}
\end{eqnarray}
The IR divergences (both soft and collinear) of the integrals
are contained in the first
term within the square brackets, while the second term is finite.
The function $G^K(s,t;{\varepsilon}_{IR};\{m_i^2\})$ is represented
by a sum of powerlike terms, it depends on ${\varepsilon}_{IR}$
and is finite in the 
${\varepsilon}_{IR} \rightarrow 0$ limit.
As for the function $H^K(s,t;\{m_i^2\})$, it is given in terms
of dilogarithmic functions and constants. 

Comparing our results with the corresponding ones of
Ref. \cite{bern}, we find that the expressions for
$G^K(s,t;{\epsilon}_{IR};\{m_i^2\})$ are in agreement.
On the other hand, the corresponding expressions for the terms
$H^K(s,t;\{m_i^2\})$ are of different form. Proving the equivalence of
our results for the integrals $I_4^K(s,t;\{m_i^2\})$ with those
of Ref. \cite{bern} then amounts to showing that the expressions for
the terms $H^K(s,t;\{m_i^2\})$ agree numerically.
By doing this, we have arrived at the following conclusions:
First, the results are in complete agreement in the Euclidean region.
Second, for the integrals $I_4^{2mh}(s,t;m_3^2,m_4^2)$,
$I_4^{1m}(s,t; m_4^2)$, and
$I_4^{0m}(s,t)$, we have found agreement for arbitrary values of
the kinematic variables.
Third, the results for the integrals \linebreak
$I_4^{3m}(s,t;,m_2^2,m_3^2,m_4^2)$ and $I_4^{2me}(s,t;m_2^2,m_4^2)$
do not agree outside the Euclidean region.

The reason for this disagreement is that the analytical
continuation from the Euclidean to the physical region as given by Eq.
(\ref{eq:f77})
is not well defined for all terms appearing in the expressions for the
integrals $I_4^K(s,t;m_i^2)$ of Ref. \cite{bern}.
This has been pointed out in Ref. \cite{binoth}.
Thus, applying the replacements (\ref{eq:f77}), no cut will 
be hit by the powerlike
terms, logarithms, and the dilogarithms with a single ratio of the kinematical
variables. In addition to this kind of terms, the expressions for the integrals
$I_4^{3m}(s,t;,m_2^2,m_3^2,m_4^2)$ and $I_4^{2me}(s,t;m_2^2,m_4^2)$, given in
Ref. \cite{bern},
contain terms of the form
\begin{equation}
{\mbox{Li}_2}
\left (
1-\frac{m_2^2m_4^2}{st}
\right ), \label{eq:fsh}
\end{equation}
i.e., the dilogarithms of a product of ratios of the kinematic variables.
This type of term requires special care. 
In order to avoid crossing a cut,
in this case one has to make the following replacements:
\begin{eqnarray}
\lefteqn{{\mbox{Li}_2}
\left (
1-\frac{m_2^2m_4^2}{st}
\right )  \rightarrow 
{\mbox{Li}_2}
\left (
1-\frac{m_2^2+{\rm i}\epsilon}{s+{\rm i}\epsilon}
  \frac{m_4^2+{\rm i}\epsilon}{t+{\rm i}\epsilon}
\right )}
\label{eq:f79} \\
& &
+
\eta
\left (
\frac{m_2^2+{\rm i}\epsilon}{s+{\rm i}\epsilon}
,\frac{m_4^2+{\rm i}\epsilon}{t+{\rm i}\epsilon}
\right )
{\ln}
\left (1-
\frac{m_2^2+{\rm i}\epsilon}{s+{\rm i}\epsilon}
\frac{m_4^2+{\rm i}\epsilon}{t+{\rm i}\epsilon}
\right ),\nonumber
\end{eqnarray}
where the function $\eta$ is defined as
\begin{equation}
{\rm \eta} (x,y)=\ln(xy)-\ln(x)-\ln(y),\label{eq:f80}
\end{equation}
and arises from the possibility that $x$, $y$, and $xy$ are not on
the same Riemann sheet.
Choosing the principal value of the logarithm such that the cut lies along
the negative real axis, Eq. (\ref{eq:f80}) can be written as follows:
\begin{eqnarray}
{\rm \eta} (x,y)&=&
2\pi {\rm i}
\{
\theta (-{\rm Im}~x)
\theta (-{\rm Im}~y)
\theta ({\rm Im}~xy)\nonumber \\
& &{}-
\theta ({\rm Im}~x)
\theta ({\rm Im}~ y)
\theta (-{\rm Im}~xy) \}~.\label{eq:f81}
\end{eqnarray}
If the terms of the form given in (\ref{eq:fsh}) are analytically continued
in accordance with
(\ref{eq:f79}), we find that the results of Ref. \cite{bern} for the 
integrals $I^{3m}_4(s, t; m_2^2, m_3^2, m_4^2)$ and 
$I^{2me}_4(s, t; m_2^2, m_4^2)$ are numerically equivalent to the
corresponding results obtained in this paper for arbitrary values of
kinematic variables.

Having thus numerically established the equivalence of the two sets of results
for the integrals $I^{K}_4(s, t; \{ m_i^2\} )$,
our next task is to demonstrate
this equivalence analytically, i.e., explicitly. There are, in principle, two
approaches to accomplish this. The first approach consists in
applying a series of the Hill identities \cite{hill} (relating the dilogarithms
of different arguments) to the final expression for the integral $Q^{3m}$ given
by Eq. (\ref{eq:f42}), with the aim to express it in terms of 
the dilogarithms occuring in
the final expression for the integral $I^{3m}_4(s, t; \{m_i^2\})$ in Ref.
\cite{bern}. However, because of the presence of ${\rm i}\epsilon$ in the arguments of the
dilogarithms in Eq. (\ref{eq:f42}), this turns out to be extremely messy. In the
second approach, one tries to achieve the same by recalculating the integral in
(\ref{eq:f38}) with an appropriate change of the integration variable. After a
lengthy calculation, details of which are presented in Appendix A, we have
been able to show that
\begin{eqnarray}
\lefteqn{Q^{3m}=
-\mbox{Li}_2
\left ( 1-
\frac{m_2^2+{\rm i}\epsilon}{s+{\rm i}\epsilon}
\right )
-\mbox{Li}_2
\left ( 1-
\frac{m_2^2+{\rm i}\epsilon}{t+{\rm i}\epsilon}
\right )}\nonumber \\
& &{}-\mbox{Li}_2
\left ( 1-
\frac{m_4^2+{\rm i}\epsilon}{s+{\rm i}\epsilon}
\right )
-\mbox{Li}_2
\left ( 1-
\frac{m_4^2+{\rm i}\epsilon}{t+{\rm i}\epsilon}
\right )\nonumber \\ & &{}+\mbox{Li}_2
\left ( 1-
\frac{m_4^2+{\rm i}\epsilon}{s+{\rm i}\epsilon}
\frac{m_2^2+{\rm i}\epsilon}{t+{\rm i}\epsilon}
\right )
-
\frac{1}{2}
\ln^2
\left (
\frac{s+{\rm i}\epsilon}{t+{\rm i}\epsilon}
\right )\nonumber \\
& & {}+
\eta
\left (
\frac{m_4^2+{\rm i}\epsilon}{s+{\rm i}\epsilon},
\frac{m_2^2+{\rm i}\epsilon}{t+{\rm i}\epsilon}
\right )
\ln
\left ( 1-
\frac{m_4^2+{\rm i}\epsilon}{s+{\rm i}\epsilon}
\frac{m_2^2+{\rm i}\epsilon}{t+{\rm i}\epsilon}
\right
)\nonumber \\
& &+\mathcal{O}({\varepsilon}_{IR}).\label{eq:fiv}
\end{eqnarray}
Although different in form, one can readily prove that this expression for
$Q^{3m}$ is numerically equivalent to that given in Eq. (\ref{eq:f42}) for
arbitrary values of kinematic variables. All
other integrals $Q^K, K\in \{ 2mh, 2me, 1m, 0m \}$ can be derived by taking appropriate zero$-$mass limits.
Combining the expressions thus obtained for $Q^K$ with expressions 
(\ref{eq:f65}), (\ref{eq:f66}), 
and
(\ref{eq:f68}) for $P^K$, we arrive at the final expressions for the integrals 
$I^{K}_4(s, t; \{ m_i^2\} )$ which are in agreement with those 
of Ref. \cite{bern} if provided the latter are correctly analytically
continued outside the Euclidean region. 

\section{Conclusion}
\label{sec:6}

Using the Feynman parameter method, we have calculated 
in an elegant manner a set
of one$-$loop box scalar integrals with massless internal lines,
but containing 0, 1, 2, or 3 nonzero external masses. To treat
IR divergences (both soft and collinear), the dimensional
regularization method has been employed. 
We have kept the causal i$\epsilon$ systematically throughout the calculation.
Consequently, 
the results for these integrals, 
which appear in the process of evaluating one$-$loop $(N\ge 5)-$point
integrals and in subdiagrams in QCD loop calculations, have been 
obtained for arbitrary values of the kinematic variables and represent the
extension of the results of Ref. \cite{bern} outside the Euclidean region.

\begin{acknowledgement}
This work was supported by the Ministry of Science and Technology
of the Republic of Croatia under Contract No. 00980102.
\end{acknowledgement}

%\appendix
\section*{Appendix A}
\label{sec:7}

In this Appendix we analytically demonstrate that for Euclidean kinematics
our results for the intgerals $I_4^K(s,t;\{m_i^2\})$
are in agreement with the corresponding results
obtained in Ref. \cite{bern}.
As a byproduct, we prove that the correct analytical continuation for the
terms of the form given in (\ref{eq:fsh}), which appear in the expressions
for the integrals $I_4^{3m}(s, t; m_2^2, m_3^2, m_4^2)$ and
$I_4^{2me}(s, t; m_2^2, m_4^2)$ of Ref. \cite{bern}, is given by Eq.
(\ref{eq:f79}).

To accomplish that, we return to the integral $Q^{3m}$ given in Eq.
(\ref{eq:f38}),
in which, for convenience,  we replace the integration variable $z$ by $y$.
Passing to the new integration variable given by
\begin{equation}
z=1-
\frac
{y(t-m_2^2)-t-{\rm i}\epsilon}
{y(m_4^2-s)-m_4^2-{\rm i}\epsilon}\;,\label{eq:f82}
\end{equation}
the integral $Q^{3m}$ becomes a line integral
\begin{eqnarray}
Q^{3m}&=&
\int_{\displaystyle {z_1}}^{\displaystyle {z_2}}
\frac{{\rm d}z}{z}
\frac{s+t-m_2^2-m_4^2}{z(m_4^2-s)+s+t-m_2^2-m_4^2}
\;\ln(1-z)\nonumber \\
& &{}+\mathcal{O}({\varepsilon}_{IR})~,\label{eq:f83}
\end{eqnarray}
with the path of integration followed from
$z_1$ to $z_2$, where
\begin{equation}
z_1=1-\frac{t+{\rm i}\epsilon}{m_4^2+{\rm i}\epsilon},~~~~
z_2=1-\frac{m_2^2+{\rm i}\epsilon}{s+{\rm i}\epsilon}\;\cdot \label{eq:f84}
\end{equation}
Applying the partial fraction decomposition (\ref{eq:f23}), this integral can be
represented as
\begin{equation}
Q^{3m}~=~I~+~J~+\mathcal{O}({\varepsilon}_{IR})~,\label{eq:f85}
\end{equation}
where
\begin{equation}
I=
\int_{\displaystyle {z_1}}^{\displaystyle {z_2}}
\frac{{\rm d}z}{z}\ln(1-z)\label{eq:f86}
\end{equation}
and
\begin{equation}
J=
-\int_{\displaystyle {z_1}}^{\displaystyle {z_2}}
{{\rm d}z}
\frac {m_4^2-s}
{z (m_4^2-s)+s+t-m_2^2-m_4^2}\ln(1-z). \label{eq:f87}
\end{equation}
Let us now consider these two integrals in turn.

The integrand in (\ref{eq:f86}) has a first$-$order pole at $z=0$
and the logarithmic branch cut extending from 1 to $\infty$.
It follows from Eq. (\ref{eq:f84}) that, depending on the values of the parameters
$s,t,m_2^2,m_4^2$, it might happen that the line connecting
the points $z_1$ and $z_2$ crosses the real axis.
In this case, as it can be seen from Eq. (\ref{eq:f82}), the crossing occurs at
the point $z=0$ $-$ the pole of the integrand.
Observe, however, that the residue of the integrand at $z=0$ is zero.
A consequence of this is that, regardless of whether the line
connecting $z_1$ and $z_2$ crosses the real axis or not, we are allowed
to assume that it passes through the point $z=0$. 
Consequently, the line
between $z_1$ and $z_2$ can be decomposed into two segments:
one connecting $z_1$ and 0, and the other connecting
$0$ and $z_2$.
In view of this, the integral $I$
can be rewritten in the form
\begin{equation}
I=
\int_0^{\displaystyle {z_2}}
\frac{{\rm d}z}{z}
\;\ln(1-z)
-\int_0^{\displaystyle {z_1}}
\frac{{\rm d}z}{z}
\;\ln(1-z)~.\label{eq:f88}
\end{equation}
By changing the integration variables
$z\rightarrow z\, z_2$ and $z\rightarrow z\, z_1$ in the first and
second integral, respectively, and taking formula (\ref{eq:f41}) 
into account,
we obtain the result
\begin{equation}
I=
-
\mbox{Li}_2
\left (1-
\frac{m_2^2+{\rm i}\epsilon}{s+{\rm i}\epsilon}
\right )
+
\mbox{Li}_2
\left (1-
\frac{t+{\rm i}\epsilon}{m_4^2+{\rm i}\epsilon}
\right )\;\cdot\label{eq:f89}
\end{equation}
Applying the transformation 
\begin{equation}
\mbox{Li}_2
\left (1-
\frac{1}{y}
\right )
=-
\mbox{Li}_2(1-y)
-\frac{1}{2}\;
\ln^2y~\label{eq:f90}
\end{equation}
to the second term on the right$-$hand side in (\ref{eq:f89}), the final expression
for the integral $I$ is found to be
\begin{eqnarray}
I&=&
-
\mbox{Li}_2
\left (1-
\frac{m_2^2+{\rm i}\epsilon}{s+{\rm i}\epsilon}
\right )
-
\mbox{Li}_2
\left (1-
\frac{m_4^2+{\rm i}\epsilon}{t+{\rm i}\epsilon}
\right )\nonumber \\
& &{}-\frac{1}{2}\;
\ln^2
\left (
\frac{m_4^2+{\rm i}\epsilon}{t+{\rm i}\epsilon}
\right )~\cdot \label{eq:f91}
\end{eqnarray}
Turning now to the integral $J$ in (\ref{eq:f87}),
we switch back to the old integration variable as a result of which
the integral takes the form
\begin{equation}
J=J_1+J_2+J_3~,\label{eq:f92}
\end{equation}
where
\begin{eqnarray}
J_1&=&-
\int_0^1 {\rm d}y
\frac{m_4^2-s}
{y\,(m_4^2-s)-
m_4^2-{\rm i}\epsilon}\nonumber \\
& &\times \ln\;
[y(m_4^2-s)-m_4^2-{\rm i}\epsilon]\;,
\label{eq:f93} \\
J_2&=&
\int_0^1 {\rm d}y
\frac{m_4^2-s}
{y\,(m_4^2-s)-
m_4^2-{\rm i}\epsilon}
\ln\;
(-t-{\rm i}\epsilon)\;,\label{eq:f94}
\end{eqnarray}
and
\begin{equation}
J_3=
\int_0^1 {\rm d}y
\frac{m_4^2-s}
{y\,(m_4^2-s)-
m_4^2-{\rm i}\epsilon}
\ln\;
\left (
1-y\frac{t-m_2^2}{t+{\rm i}\epsilon}
\right )\;\cdot \label{eq:f95}
\end{equation}
The integrals $J_1$ and $J_2$ are elementary, and are readily
evaluated. The results are
\begin{eqnarray}
J_1&=&
-\frac{1}{2}
\;\left [
\;\ln^2(-s-{\rm i}\epsilon)-
\ln^2(-m_4^2-{\rm i}\epsilon)\;
\right ]\;,\label{eq:f96}
\\
J_2&=&
\ln\;
(-t-{\rm i}\epsilon)\;
\ln\;
\left (
\frac{s+{\rm i}\epsilon}{m_4^2+{\rm i}\epsilon}
\right )~\cdot \label{eq:f97}
\end{eqnarray}
In order to evaluate the integral $J_3$,
we proceed by adding and subtracting the following expression:
\[
\int_0^1
\frac {{\rm d}y}{y} 
\ln
\left (
1-y\frac{t-m_2^2}{t+{\rm i}\epsilon}
\right )
+
\int_0^1
\frac{{\rm d}y}{y} 
\ln(1-y)\]
\[+
\int_0^1{\rm d}y
\frac{1}
{y-
{\displaystyle
\frac
{m_4^2+{\rm i}\epsilon}{m_4^2-s}}}
\ln(1-y)~.
\]
After a simple algebraic reduction, the integral $J_3$ can be
represented in the form
\begin{equation}
J_3=J_{3,1}+J_{3,2}\;,\label{eq:f98}
\end{equation}
where
\begin{eqnarray}
\lefteqn{J_{3,1}=
\int_0^1
\frac {{\rm d}y}{y} 
\ln
\left (
1-y\frac{t-m_2^2}{t+{\rm i}\epsilon}
\right )
-
\int_0^1
\frac{{\rm d}y}{y} 
\ln(1-y)}\nonumber \\
& &{}+
\int_0^1{\rm d}y
\frac{1}
{y-
{\displaystyle
\frac
{m_4^2+{\rm i}\epsilon}{m_4^2-s}}}
\ln(1-y)
\nonumber \\
&=&
-\mbox{Li}_2
\left (1-\frac{m_2^2+{\rm i}\epsilon}{t+{\rm i}\epsilon} \right )
+\frac{{\pi}^2}{6}
-\mbox{Li}_2
\left (1-\frac{m_4^2+{\rm i}\epsilon}{s+{\rm i}\epsilon} \right
)\;,\label{eq:f99}
\end{eqnarray}
and
\begin{equation}
J_{3,2}=
-\int_0^1{\rm d}y
\left [
\frac{1}
{y-
{\displaystyle
\frac
{m_4^2+{\rm i}\epsilon}{m_4^2-s}}}
-\frac{1}{y}
\right ]
\ln
\left (
\frac
{1-y}
{1-y\displaystyle {\frac{t-m_2^2}{t+{\rm i}\epsilon}}}
\right )\;.\label{eq:f100}
\end{equation}
Next, consider the integral $J_{3,2}$.
Introducing a new integration variable
\begin{equation}
z=
\frac
{1-y}
{1-y\displaystyle {\frac{t-m_2^2}{t+{\rm i}\epsilon}}}~,\label{eq:f101}
\end{equation}
and the notation
\begin{equation}
a=
\frac
{(m_2^2+{\rm i}\epsilon)(m_4^2+{\rm i}\epsilon)}
{(s+{\rm i}\epsilon)(t+{\rm i}\epsilon)}~,\label{eq:f102}
\end{equation}
the integral can be cast into the form
\begin{equation}
J_{3,2}=
\int_0^1 \frac{{\rm d}z}{z}
\ln(1-z)
+
\int_0^1 {\rm d}z
\frac
{1}
{z-\displaystyle
{\frac{1}{1-a}}}
~\ln z~.\label{eq:f103}
\end{equation}
Notice that, in accordance with (\ref{eq:f101}), the integral $J_{3,2}$ is
given by Eq. (\ref{eq:f103}) as a line integral
in the complex $z-$plane with the integration path having only
the end$-$po\-ints on the real axis at
$z=0$ and $z=1$.
Adding and subtracting  the integral
of the form
\[
\int_0^1 {\rm d}z
\frac
{1}
{z-\displaystyle
{\frac{1}{1-a}}}
~\ln(1-a)
\]
in the second integral in (\ref{eq:f103}) allows us to write
the integral $J_{3,2}$ as
\begin{eqnarray}
J_{3,2}&=&
\int_0^1 \!\!\frac{{\rm d}z}{z}
\ln(1-z)
+
\int_0^1 \!\!{\rm d}z
\frac
{1}
{z-\displaystyle
{\frac{1}{1-a}}}
[\ln z +
\ln(1-a)]\nonumber \\
& &{}-
\int_0^1 {\rm d}z
\frac
{1}
{z-\displaystyle
{\frac{1}{1-a}}}
~\ln(1-a)~.\label{eq:f104}
\end{eqnarray}
It should be observed that the residue of the first integral 
on the right$-$hand side
in (\ref{eq:f101}) at the pole $z=0$ is equal to zero. The same is true
for the second integral at the pole $z=1/(1-a)$. Therefore, the
integration path in both of these integrals can be taken to follow
the real axis from 0 to 1.
After evaluating the first two integrals in (\ref{eq:f104}) and
passing to the old integration variable in the third integral, we arrive
at the following expression for $J_{3,2}$:
\begin{eqnarray}
\lefteqn{J_{3,2}=
-\frac {{\pi}^2}{6}+
\mbox{Li}_2(1-a)
+\ln(a)
\ln(1-a)+\ln(1-a)}
\nonumber \\
& &
{}\!\!\!\!\times \!
\int_0^1 \!\!{\rm d}y
\frac{st-m_2^2m_4^2 +{\rm i}\epsilon (s+t-m_2^2-m_4^2)}
{[y(m_4^2-s)-m_4^2-{\rm i}\epsilon][y(t-m_2^2)-t-{\rm i}\epsilon]}
\;.\label{eq:f105}
\end{eqnarray}
Carrying out the remaining integration,
we obtain the final expression 
\begin{equation}
J_{3,2}=
-\frac {{\pi}^2}{6}+
\mbox{Li}_2(1-a)
+
\eta
\left (
\frac{m_4^2+{\rm i}\epsilon}{s+{\rm i}\epsilon}~,
\frac{m_2^2+{\rm i}\epsilon}{t+{\rm i}\epsilon}~
\right ) \ln(1-a)~,\label{eq:f106}
\end{equation}
where the function $\eta (x, y)$ is defined by Eq. (\ref{eq:f80}).
\noindent
Now, substituting (\ref{eq:f99}) and (\ref{eq:f106}) into (\ref{eq:f98}), 
we find 
the integral $J_3$ to be given by 
\begin{eqnarray}
J_3&=&
-\mbox{Li}_2
\left ( 1-
\frac{m_2^2+{\rm i}\epsilon}{t+{\rm i}\epsilon}
\right )
-\mbox{Li}_2
\left ( 1-
\frac{m_4^2+{\rm i}\epsilon}{s+{\rm i}\epsilon}
\right ) \nonumber\\
& &{}+\mbox{Li}_2
\left ( 1-
\frac{m_4^2+{\rm i}\epsilon}{s+{\rm i}\epsilon}
\frac{m_2^2+{\rm i}\epsilon}{t+{\rm i}\epsilon}
\right )\nonumber \\
& &
{}+
\eta
\left (
\frac{m_4^2+{\rm i}\epsilon}{s+{\rm i}\epsilon},
\frac{m_2^2+{\rm i}\epsilon}{t+{\rm i}\epsilon}
\right )
\ln
\left ( 1-
\frac{m_4^2+{\rm i}\epsilon}{s+{\rm i}\epsilon}
\frac{m_2^2+{\rm i}\epsilon}{t+{\rm i}\epsilon}
\right )\cdot \nonumber \\
& &\label{eq:f107a}
\end{eqnarray}
Finally, having obtained all the necessary ingredients,
we now combine them to obtain (\ref{eq:fiv}), which is the desired result.

\section*{Appendix B}
\label{sec:8}

By expanding the powerlike terms appearing in Eqs.
(\ref{eq:f71})$-$ (\ref{eq:f75}),
and the prefactor $r_{\Gamma}$ defined by (\ref{eq:f70}),
we find that the integrals under consideration
can be written in the generic form
\begin{eqnarray}
I_{\displaystyle 4}^K&=&\frac{{\rm i}}{(4\pi )^2}\,
g^K
\left( \,
\frac{A^K}{\varepsilon_{IR}^2}+\frac{B^K}{\varepsilon_{IR}}
+C_1^K\,
+C_2^K\,\right)
+\mathcal{O}({\varepsilon}_{IR}),\nonumber \\
& &
K\in \{3m,2mh,2me,1m,0m\}\label{eq:f109}
\end{eqnarray}
The functions $g^K$ appearing above are defined by (\ref{eq:f29}).
For convenience, the finite parts have been decomposed into two terms where
the term $C_1^K$ originates from the expansion
of the product of the $r_{\Gamma}$ prefactor with the powerlike terms.
 
The double$-$pole parts $A^K$, the single$-$pole parts $B^{K}$, as well
as the finite parts $C_1^K$ and $C_2^K$
of the individual integrals are listed below. 
\begin{eqnarray}
A^{3m}&=& 0~,\nonumber\\
B^{3m}&=& \ln\!\left( \frac{s+{\rm i}\epsilon}
{m_2^2+{\rm i}\epsilon}\right)
+\ln\!\left(
\frac{t+{\rm i}\epsilon}{m_4^2+{\rm i}\epsilon}\right)~,\nonumber\\
C_1^{3m}&=& \left[{\gamma}_E+\ln\!\left( \frac{-s-{\rm i}\epsilon}
{4\pi \mu^2}\right)\right]^2\!\!
+\!\left[{\gamma}_E+\ln\!\left( \frac{-t-{\rm i}\epsilon}{4\pi
\mu^2}\right)\right]^2\nonumber\\
&-&\left[{\gamma}_E+\ln\!\left( \frac{-m_2^2-{\rm i}\epsilon}{4\pi
\mu^2}\right)\right]^2\!\!
-\!\left[{\gamma}_E+\ln\!\left( \frac{-m_3^2-{\rm i}\epsilon}{4\pi
\mu^2}\right)\right]^2\nonumber\\
&-&\left[{\gamma}_E+\ln\!\left( \frac{-m_4^2-{\rm i}\epsilon}{4\pi
\mu^2}\right)\right]^2\nonumber\\
&+&\frac{1}{2}\,\left[{\gamma}_E+\ln\!\left( \frac{-m_3^2-{\rm i}\epsilon}
{4\pi \mu^2}\right)
+\ln\!\left(
\frac{m_4^2+{\rm i}\epsilon}
{s+{\rm i}\epsilon}\right)\right]^2\nonumber\\
&+&\frac{1}{2}\,\left[{\gamma}_E+\ln\!
\left( \frac{-m_3^2-{\rm i}\epsilon}
{4\pi \mu^2}\right)
+\ln\!\left( \frac{m_2^2+{\rm i}\epsilon}
{t+{\rm i}\epsilon}\right)\right]^2~,\nonumber\\
C_2^{3m}&=& 
2\, \mbox{Li}_{2}\left(\, 1-\frac{m_{2}^{2}+{\rm i}\epsilon }{t+{\rm i}\epsilon}\,
\right)+2\,\mbox{Li}_{2}\left(\, 1-\frac{m_{4}^{2}+{\rm i}\epsilon}
{s+{\rm i}\epsilon}\, \right)\nonumber\\
& &\hspace{-1cm}{}+2\, \mbox{Li}_{2}\Big[\, 1-(s+{\rm i}\epsilon)\, f^{3m}\,\Big]
\!+2\, \mbox{Li}_{2}\Big[\,
1-(t+{\rm i}\epsilon)\, f^{3m}\, \Big]\nonumber\\
& &\hspace{-1cm}{}-2\, \mbox{Li}_{2}\Big[\, 1-(m_{2}^{2}+{\rm i}\epsilon)\, f^{3m}\, \Big]
\!-2\,
\mbox{Li}_{2}\Big[\,
1-(m_{4}^{2}+{\rm i}\epsilon)\,
f^{3m}\,\Big].\nonumber \\
& & 
\end{eqnarray}
\begin{eqnarray}
A^{2mh}&=& 1~,\nonumber\\
B^{2mh}&=& {\gamma}_E+\ln\!\left( \frac{-t-{\rm i}\epsilon}
{4\pi \mu^2}\right)
+\ln\!\left( \frac{s+{\rm i}\epsilon}
{m_4^2+{\rm i}\epsilon}\right)\nonumber\\
&+&\ln\!\left(
\frac{t+{\rm i}\epsilon}{m_3^2+{\rm i}\epsilon}\right)~,\nonumber\\
C_1^{2mh}&=& \left[{\gamma}_E+\ln\!\left( \frac{-s-{\rm i}\epsilon}
{4\pi \mu^2}\right)\right]^2
+\left[{\gamma}_E+\ln\!\left( \frac{-t-{\rm i}\epsilon}
{4\pi \mu^2}\right)\right]^2\nonumber\\
& &\hspace{-1cm}{}-\left[{\gamma}_E+\ln\!\left( \frac{-m_3^2-{\rm i}\epsilon}{4\pi
\mu^2}\right)\right]^2
-\left[{\gamma}_E+\ln\!\left( \frac{-m_4^2-{\rm i}\epsilon}
{4\pi \mu^2}\right)\right]^2\nonumber \\
& &\hspace{-1cm}{}+\frac{1}{2}\,\left[{\gamma}_E+\ln\!\left( \frac{-m_3^2-{\rm i}\epsilon}
{4\pi \mu^2}\right)
+\ln\!\left( \frac{m_4^2+{\rm i}\epsilon}
{s+{\rm i}\epsilon}\right)\right]^2
-\frac{\pi^2}{12}~,\nonumber\\
C_2^{2mh}&=&
2\,\mbox{Li}_{2}\left(\, 1-\frac{m_{4}^{2}+{\rm i}\epsilon }{s+{\rm i}\epsilon}\,
\right)-2\,\mbox{Li}_{2}\left(\, 1-\frac{m_{3}^{2}+{\rm i}\epsilon}
{t+{\rm i}\epsilon}\, \right)\nonumber\\
& &\hspace{-1cm}
{}+2\, \mbox{Li}_{2}\Big[\, 1-(s+{\rm i}\epsilon)\, f^{2mh}\, \Big]
+2\, \mbox{Li}_{2}\Big[\,
1-(t+{\rm i}\epsilon)\, f^{2mh}\,  \Big]\nonumber\\
& &\hspace{-1cm}
{}-2\, \mbox{Li}_{2}\Big[\, 1-(m_{4}^{2}+{\rm i}\epsilon)\, f^{2mh}\, \Big].
\end{eqnarray}
\begin{eqnarray}
A^{2me}&=& 0~,\nonumber\\
B^{2me}&=& 2\,\ln\!\left( \frac{s+{\rm i}\epsilon}
{m_2^2+{\rm i}\epsilon}\right)
+2\,\ln\!\left(
\frac{t+{\rm i}\epsilon}{m_4^2+{\rm i}\epsilon}\right)~,\nonumber\\
C_1^{2me}&=&\left[{\gamma}_E+\ln\!\left( \frac{-s-{\rm i}\epsilon}
{4\pi \mu^2}\right)\right]^2
+\left[{\gamma}_E+\ln\!\left( \frac{-t-{\rm i}\epsilon}
{4\pi \mu^2}\right)\right]^2\nonumber\\
& &\hspace{-1cm}{}-\left[{\gamma}_E+\ln\!\left( \frac{-m_2^2-{\rm i}\epsilon}{4\pi
\mu^2}\right)\right]^2
-\left[{\gamma}_E+\ln\!\left( \frac{-m_4^2-{\rm i}\epsilon}
{4\pi \mu^2}\right)\right]^2~, \nonumber\\
C_2^{2me}&=&
2\, \mbox{Li}_{2}\Big[\, 1-(s+{\rm i}\epsilon)\, f^{2me}\, \Big]
\!\!+\!2\, \mbox{Li}_{2}\Big[\,
1-(t+{\rm i}\epsilon)\, f^{2me}\, \Big]\nonumber\\
& &\hspace{-1.2cm}{}-\!2\, \mbox{Li}_{2}\Big[\, 1-(m_{2}^{2}+{\rm i}\epsilon)\, f^{2me}\, \Big]
\!\!-\!2\, \mbox{Li}_{2}\Big[\,
1-(m_{4}^{2}+{\rm i}\epsilon)\, f^{2me}\,
\Big].\nonumber
\\& &
\end{eqnarray} 
\begin{eqnarray}
A^{1m}&=& 2~,\nonumber\\
B^{1m}&=& 2\,{\gamma}_E+2\,\ln\!\left( \frac{-s-{\rm i}\epsilon}
{4\pi \mu^2}\right)
+2\,\ln\!\left( \frac{-t-{\rm i}\epsilon}{4\pi \mu^2}\right)
\nonumber \\
&-&2\,\ln\!\left( \frac{-m_4^2-{\rm i}\epsilon}
{4\pi \mu^2}\right)~,\nonumber\\
C_1^{1m}&=& \left[{\gamma}_E+\ln\!\left( \frac{-s-{\rm i}\epsilon}
{4\pi \mu^2}\right)\right]^2
+\left[{\gamma}_E+\ln\!\left( \frac{-t-{\rm i}\epsilon}
{4\pi \mu^2}\right)\right]^2\nonumber \\
&-&\left[{\gamma}_E+\ln\!\left( \frac{-m_4^2-{\rm i}\epsilon}
{4\pi \mu^2}\right)\right]^2-\frac{\pi^2}{6}~,\nonumber\\
C_2^{1m}&=& 
2\,\mbox{Li}_{2}\Big[\, 1-(s+{\rm i}\epsilon)\, f^{1m} \,\Big]
+2\,\mbox{Li}_{2}\Big[\,
1-(t+{\rm i}\epsilon)\,f^{1m} \, \Big]\nonumber\\
&-&2\,\mbox{Li}_{2}\Big[\, 1-(m_{4}^{2}+{\rm i}\epsilon)\,f^{1m}\, \Big]
-\frac{\pi^{2}}{3}~.
\end{eqnarray}
\begin{eqnarray}
A^{0m}&=& 4~,\nonumber\\
B^{0m}&=& 4\,{\gamma}_E+2\,\ln\!\left( \frac{-s-{\rm i}\epsilon}
{4\pi \mu^2}\right)
+2\,\ln\!\left( \frac{-t-{\rm i}\epsilon}{4\pi \mu^2}\right)~,\nonumber\\
C_1^{0m}&=& \!\left[{\gamma}_E+\ln\!\left( \frac{-s-{\rm i}\epsilon}
{4\pi \mu^2}\right)\right]^2
\!\!\!+\left[{\gamma}_E+\ln\!\left( \frac{-t-{\rm i}\epsilon}
{4\pi \mu^2}\right)\right]^2
\!\!\!-\frac{\pi^2}{3},\nonumber\\
C_2^{0m}&=&
2\,\mbox{Li}_{2}\Big[\, 1-(s+{\rm i}\epsilon)\,f^{0m}\, \Big]
+2\,\mbox{Li}_{2}\Big[\,
1-(t+{\rm i}\epsilon)\,f^{0m} \, \Big]\nonumber
\\
&-&2\, \frac{\pi^{2}}{3}~. 
\end{eqnarray}
In the above expressions, ${\gamma}_E$ = 0.5722 is
the Euler constant.

The pole and finite parts for the integrals
$I_4^K(s,t;{m_i^2)}$ obtained in Ref. \cite{bern}, but with the correct
analytical continuation outside the Euclidean region for the
integrals $I_4^{3m}(s,t;m_2^2,m_3^2,m_4^2)$ and $I_4^{2me}(s,t;m_2^2,m_4^2)$,
can be obtained simply by replacing the terms $C_2^K(s,t;m_i^2)$ given
above by the corresponding values listed below:
\begin{eqnarray}
C_2^{3m}&=&
-\ln^2
\left (
\frac
{s+{\rm i}\epsilon}{t+{\rm i}\epsilon}
\right )\nonumber\\
& &\hspace{-1cm}{}-2\;\mbox{Li}_2
\left (
1-
\frac
{m_2^2+{\rm i}\epsilon}{s+{\rm i}\epsilon}
\right )
-2\;\mbox{Li}_2
\left (
1-
\frac
{m_4^2+{\rm i}\epsilon}{t+{\rm i}\epsilon}
\right )\nonumber\\
& &\hspace{-1cm}{}+2\;\mbox{Li}_2
\left (
1-\frac{m_2^2+{\rm i}\epsilon}{s+{\rm i}\epsilon}
  \frac{m_4^2+{\rm i}\epsilon}{t+{\rm i}\epsilon}
\right )\nonumber\\
& &\hspace{-1cm}{}+
2\;\eta
\left (
\frac{m_2^2+{\rm i}\epsilon}{s+{\rm i}\epsilon},
\frac{m_4^2+{\rm i}\epsilon}{t+{\rm i}\epsilon}
\right )
\ln
\left (1-
\frac{m_2^2+{\rm i}\epsilon}{s+{\rm i}\epsilon}
\frac{m_4^2+{\rm i}\epsilon}{t+{\rm i}\epsilon}
\right )~,\nonumber \\& &
\end{eqnarray}
\begin{eqnarray}
C_2^{2mh}&=&
-\ln^2
\left (
\frac
{s+{\rm i}\epsilon}{t+{\rm i}\epsilon}
\right )\nonumber\\
& &\hspace{-1cm}{}-2\;\mbox{Li}_2
\left (
1-
\frac
{m_3^2+{\rm i}\epsilon}{t+{\rm i}\epsilon}
\right )
-2\;\mbox{Li}_2
\left (
1-
\frac
{m_4^2+{\rm i}\epsilon}{t+{\rm i}\epsilon}
\right )~,\nonumber \\ & &
\end{eqnarray}
\begin{eqnarray}
C_2^{2me}&=&
-\ln^2
\left (
\frac
{s+{\rm i}\epsilon}{t+{\rm i}\epsilon}
\right )\nonumber\\
& &\hspace{-1cm}{}-2\;\mbox{Li}_2
\left (
1-
\frac
{m_2^2+{\rm i}\epsilon}{s+{\rm i}\epsilon}
\right )
-2\;\mbox{Li}_2
\left (
1-
\frac
{m_2^2+{\rm i}\epsilon}{t+{\rm i}\epsilon}
\right )\nonumber\\
& &\hspace{-1cm}{}-2\;\mbox{Li}_2
\left (
1-
\frac
{m_4^2+{\rm i}\epsilon}{s+{\rm i}\epsilon}
\right )
-2\;\mbox{Li}_2
\left (
1-
\frac
{m_4^2+{\rm i}\epsilon}{t+{\rm i}\epsilon}
\right )\nonumber\\
& &\hspace{-1cm}{}+2\;\mbox{Li}_2
\left (
1-\frac{m_2^2+{\rm i}\epsilon}{s+{\rm i}\epsilon}
  \frac{m_4^2+{\rm i}\epsilon}{t+{\rm i}\epsilon}
\right )\nonumber\\
& &\hspace{-1cm}
{}+2\;\eta
\left (
\frac{m_2^2+{\rm i}\epsilon}{s+{\rm i}\epsilon},
\frac{m_4^2+{\rm i}\epsilon}{t+{\rm i}\epsilon}
\right )
\ln
\left (1-
\frac{m_2^2+{\rm i}\epsilon}{s+{\rm i}\epsilon}
\frac{m_4^2+{\rm i}\epsilon}{t+{\rm i}\epsilon}
\right ),\nonumber \\ & &
\end{eqnarray} 
\begin{eqnarray}
C_2^{1m}&=&
-\ln^2
\left (
\frac
{s+{\rm i}\epsilon}{t+{\rm i}\epsilon}
\right )
-\frac
{{\pi}^2}{3}\nonumber\\
& &\hspace{-1cm}{}-2\;\mbox{Li}_2
\left (
1-
\frac
{m_4^2+{\rm i}\epsilon}{s+{\rm i}\epsilon}
\right )
-2\;\mbox{Li}_2
\left (
1-
\frac
{m_4^2+{\rm i}\epsilon}{t+{\rm i}\epsilon}
\right ),
\\ & &\nonumber\\
C_2^{0m}&=&
-\ln^2
\left (
\frac
{s+{\rm i}\epsilon}{t+{\rm i}\epsilon}
\right )
-{\pi}^2~.
\end{eqnarray}
As it is readily seen from the above expressions,
the integrals $I_4^K(s,t;\{m_i^2\})$ are real in the Euclidean region .
Outside of this region, however, the integrals acquire an imaginary part.
Being given in terms of logarithms and dilogarithms, their imaginary parts
can be easily determined.

Thus, with the usual definition of the logarithms on the branch cut
$-\infty < {\rm z} \le 0$, one has
\begin{equation}
\ln(y\pm i\epsilon)=\ln|y|\pm {\rm i}\pi \theta (-y)~.
\end{equation}
Next, as it is seen from Eq. (\ref{eq:f41}), the function ${\rm {Li_2}}(y)$
develops an imaginary part for $y\ge 1$, and 
\begin{equation}
{\rm {Re~ Li_2}} (y\pm {\rm i}\epsilon )=
- {\rm Li_2}
\left (\frac{1}{y}\right )-\frac{1}{2}\ln^2y+
\frac{{\pi}^2}{3}\:,
\end{equation}
\begin{equation}
{\rm Im~Li_2} (y\pm {\rm i}\epsilon )=\pm \pi \ln~y.
\end{equation}
 
% BibTeX users please use
% \bibliographystyle{}
% \bibliography{}

\end{document}